# A comprehensive multiphonon spectral analysis in $MoS_2$

Tsachi Livneh[1*#] and Jonathan E. Spanier[2]

[1] *Department of Physics, Nuclear Research Center, Negev, P.O. Box 9001, Beer- Sheva, 84190, Israel.*
[2] *Department of Materials Science & Engineering, and Department of Physics, Drexel University, 3141 Chestnut St., Philadelphia, PA, 19104, USA.*

ABSTRACT

We present a comprehensive multiphonon Raman and complementary infrared analysis for bulk and monolayer $MoS_2$. For the bulk the analysis consists of symmetry assignment from which we obtain a broad set of allowed second-order transitions at the high symmetry M, K and Γ Brillouin zone (BZ) points. The attribution of about 80 transitions of up to fifth order processes are proposed in the low temperature (95 K) resonant Raman spectrum measured with excitation energy of 1.96 eV, which is slightly shifted in energy from the *A* exciton. We propose that the main contributions come from four phonons: $A_{1g}$ (M), $E^1_{2g}$ ($M_2$), $E^2_{2g}$ ($M_1$) (TA' (M)) and $E^2_{2g}$ ($M_2$) (LA' (M)). The last three are single degenerate phonons at M with an origin of the $E^1_{2g}$ (Γ) and $E^2_{2g}$ (Γ) phonons. Among the four phonons, we identify in the resonant Raman spectra all (but one) of the second-order overtones, combination and difference-bands and many of the third order bands. Consistent with the expectation that at the M point only combinations with the same inversion symmetry (*g* or *u*) are Raman-allowed, the contribution of combinations with the LA(M) mode can not be considered with the above four phonons. Although minor, contributions from K point and possibly Γ-point phonons are also evident.
The '2LA band', measured at ~ 460 $cm^{-1}$ is reassigned. Supported by the striking similarity between this band, measured under off-resonant conditions, and recently published two phonon density of states, we propose that the lower part of the band, previously attributed to 2LA(M), is due to a van Hove singularity between K and M. The higher part, previously attributed exclusively to the $A_{2u}$ (Γ) phonon, is mostly due to the LA and LA' phonons at M.
For the monolayer $MoS_2$ the second-order phonon processes from the M and Γ Brillouin zone points are also analyzed and are discussed within similar framework to that of the bulk.

# 1. Introduction

Recently, Raman scattering has been increasingly important in the study of transition- metal-layered–type-dichalcogenides [1-16]. Among those, the most investigated system is the indirect semiconductor $MoS_2$, with the *2H* hexagonal polytype ($D^4_{6h}$ space group #194), which becomes a direct band-gap semiconductor in monolayer *1H* polytype [16]. Raman scattering was employed for the various forms of bulk [1-6], Inorganic Fullerenes (IF) [7] and few layer (FL) structures down to the monolayer [8-13]. A group-theoretical analysis of the optical lattice vibrations for the bulk [1] reveals four Raman-active modes corresponding to the following symmetries with measured frequencies under ambient conditions: $E^2_{2g}$ (35 $cm^{-1}$), $E_{1g}$ (286 $cm^{-1}$), $E^1_{2g}$ (383 $cm^{-1}$), and $A_{1g}$ (408 $cm^{-1}$). In addition, there are two IR-active modes: $E_{1u}$ (384 $cm^{-1}$), $A_{2u}$ (470 $cm^{-1}$), and four silent modes: $B^2_{2g}$ (58 $cm^{-1}$) [11], $E_{2u}$ (287 $cm^{-1}$), $B_{1u}$ (403 $cm^{-1}$) and $B^1_{2g}$ (~ 475 $cm^{-1}$). In addition to observation of first-order Raman lines these and other studies showed a rich multiphonon spectrum [2, 3, 6, 7, 9, 10, 13-15] with sensitivity to excitation energy [6, 7, 9, 10, 13].

The higher order spectra of bulk $MoS_2$ have been assigned [2, 3] to be mostly constructed from second-order transitions (some of which include the longitudinal acoustic (LA) phonon at the M Brillouin zone (BZ) edge, LA(M), and BZ center Γ-point phonons). Stacy and Hodul [3] focused on the nature of the band around ~ 460 $cm^{-1}$, which is denoted hereafter as the '2LA band' and assigned it to a second-order process involving the LA(M) phonon. Frey *et al.* [7] studied this band for inorganic fullerenes (IF) and bulk $MoS_2$, suggesting that the asymmetric features of 2LA band are due to a combination of two peaks centered at room temperature at ~454 $cm^{-1}$ (denoted here as $\alpha_1$) and ~ 465 $cm^{-1}$ ($\alpha_2$). The first is assigned to the 2LA(M) and the second to a Raman-forbidden IR-allowed optical $A_{2u}$ (Γ) mode, which involves asymmetric translation of both Mo and S atoms parallel to the *c* axis [1]. It was argued that, although not allowed by Raman selection rules, under resonance conditions excitons could mediate the scattering of this phonon. The above assignment of the '2LA band' has been adopted exclusively in the literature [6-15].

Although the K-point phonons may also contribute, it was proposed that due to the fact that the 2LA(K) frequency was found (according to Ref. 17) to be higher by ~10 $cm^{-1}$ at the K point relative to the M point, their contribution is minor. This was in contrast to *2H*-$WS_2$, which was argued by Sourisseau *et al.* [17] to have a multiphonon spectrum mostly constructed from combination and difference bands with LA phonons at the K point.

Here we present a comprehensive multiphonon analysis in bulk *2H*-$MoS_2$. Supported by new evidence, we propose modified interpretations and assignments of some of the central spectral features. Significantly, our findings challenge the widely accepted one [3] according to which the majority of the observed second-order combination and difference bands from the M point are due to one of the $A_{1g}$(M), and to what is referred to as $E_{1g}$(M) and $E_{2g}$(M) phonons with LA(M) phonons, or as recently proposed by Golasa *et al.* [14, 15], with transverse acoustic (TA(M)) and/or out-of-plane transverse acoustic (ZA(M)) phonons. We complement our analysis by exploring the multiphonon spectrum of monolayer *1H*-$MoS_2$ (while leaving FL systems for later study) and anticipate that our full study will significantly advance fundamental understanding of the origin of multiphonon resonant inelastic light scattering processes in layered dichalcogenides and their application.

## 2. Experiment

Resonant Raman spectra for a bulk sample were measured in backscattering configuration using a Jobin-Yvon LabRam HR spectrometer with a He-Ne 632.8 nm laser (1.96 eV), which is slightly shifted in energy from the *A* exciton [18]. The scattered light was dispersed by a 1800 grooves/mm grating resulting in a <1 $cm^{-1}$ spectral resolution. The low temperature spectra were measured at 95 K by means of a Linkam model THMS600 continuously - cooled liquid-nitrogen stage. UV-Raman was measured with He-Cd 325 nm laser (3.81 eV).

Off-resonant Raman spectra for bulk $MoS_2$ were also measured in the backscattering configuration using a Renishaw spectrometer with a 785 nm (1.58 eV) and a 514.5 nm (2.41 eV) lasers. Raman spectra measured at 1064 nm (1.16 eV) were acquired by using Bruker FT-Raman spectrometer. Raman spectra from a monolayer of $MoS_2$ were measured at 632.8 nm (1.96 eV). The power of the laser was kept sufficiently low to avoid heating effects.

Single and few-layer $MoS_2$ films were isolated from bulk $MoS_2$ crystals by mechanical exfoliation method and placed on a silicon substrate covered by a thick $SiO_2$ layer. After confirming that the distinctive Stokes Raman spectrum of the monolayer was consistent with previously published spectra [9], the monolayer thickness was verified by topographic-height scanning probe microscopy (Asylum Research MFP-3D).

## 3. Results and discussions

### A. Symmetry mode analysis of multiphonon bulk *2H*-$MoS_2$

In hexagonal $MoS_2$ there are six atoms in the unit cell and therefore 18 branches of the phonon dispersion relation, some of which are degenerate in high symmetry directions [19]. **Table 1** presents a list of phonons in *2H*-$MoS_2$ with their symmetry assignments, divided into the phonons symmetries at Γ (Brillouin zone-center) and at M and K (zone-edge points in the (ξ00) and (ξξ0) directions in ***k*** space, respectively). For the $D^4_{6h}$ space group the eigenstates at M and K points correspond to irreducible representations of the point groups $D_{2h}$ and $D_{3h}$, respectively. In Table 1 we also show the frequencies of the various phonons. For the Γ point we show experimentally measured frequencies at 300 K. For the M and K points the listed frequencies are the proposed values that are expected at low temperatures, in accordance with DFT calculations (and will be used to assign the resonant Raman spectrum at 95 K as described in Section B). We followed Sourisseau *et al.* [17] in assigning the symmetry of the M point BZ phonons, while consulting the correlation tables between the point group of $D_{6h}$ and its subgroups [20]. In labeling the symmetry of the representations we use the Mulliken notation [21]. Major aspects of constructing Table 1 are discussed in the supporting information.

Each phonon branch is denoted with a letter (capital for Γ, Greek alphabet for M and regular for K). The phonon notation $E_{1g}(M_1)$, for example, denotes one of a doubly degenerate phonon at Γ of $E_{1g}$ symmetry, which splits and extends to M, where it has symmetry belonging to the $D_{2h}$ point group. We note that due to this splitting there are two different phonons at M ($M_1$ and $M_2$). Hence, the extensively used notation [3, 15, 16] of phonons at M that originate from doubly degenerate phonons at Γ (like $E_{1g}(M)$ and $E^1_{2g}(M)$) is not suitable. In Table 1 we also distinguish acoustic phonons as LA, TA or ZA and we denote the phonons at M originating as quasi-acoustic optical $B^2_{2g}(\Gamma)$ and $E^2_{2g}(\Gamma)$ phonons as ZA'($B^2_{2g}(M)$) and TA'($E^2_{2g}(M_1)$)+ LA'($E^2_{2g}(M_2)$), respectively. The LA' and TA' phonons will be argued to be central in the multiphonon resonant process.

At the K point all the phonons are assigned as singly degenerate, excluding two doubly degenerate modes which originate from $E_{1g}(M_1)$+$E_{2u}(M_2)$ and $E^1_{2g}(M_1)$+$E^1_{1u}(M_2)$ (see supporting information).

Key in the analysis is the correlation of experiments with symmetry-based prediction of multiphonon transitions. Observation of high-order multiphonon processes demands accurate predictions. There are few published DFT calculations [22- 24] that present the various phonon dispersions in *2H*-$MoS_2$. We extracted from those the calculated frequencies of the various M and K-point phonons. Correlating experiment with calculation, particularly for the M-point phonons, and noting that DFT estimates are for 0 K and will have at best a few % error in calculated mode energies, we introduce a tolerance of up to $\pm$ 10 $cm^{-1}$ around the low temperatures measured value to facilitate the analysis. This is due to our ability to 'fine tune' those frequencies by carefully looking at the resonant multiphonon bands, as shown below. Since some of the frequencies participate in several transitions involving high-order processes there is high sensitivity to our changes. Therefore the frequencies that we show in Table 1 are proposed and should be considered as such. Indeed, the excellent fit between the predictions and the data for the M-point phonons indicates that those are very good predictions, particularly for the M point frequencies.

We next present a derivation of the symmetries of the Raman tensors of the second-order scattering from zone-edge phonons at M and K. We have listed in Table 1 the symmetries of the individual phonons at Γ, M and K points. We next determine the irreducible representations of the binary combinations at each of these points. For the Brillouin zone-center phonons with $\boldsymbol{k} = 0$ this is accomplished by simply multiplying together the characters of the individual group operations [25a]. The determination of transitions for $\boldsymbol{k} \neq 0$ phonons is less straightforward than that for $\boldsymbol{k} = 0$, particularly for cases of non-symmorphic space groups, which are having glide planes and screw axis, e.g., space group #194, $D^4_{6h}$. Using the Bilbao crystallographic server [26] we obtained the Raman-active scattering tensors for second-order processes from phonons at M and K in the $D^4_{6h}$ space group and established the correlations between irreducible representations of the combinations of the group of a particular $\boldsymbol{k}$ and the irreducible representations of the full space group. For a binary combination to be Raman-active at the M and K points it must correlate with at least one of the three Raman-active symmetries: $A_{1g}$, $E_{2g}$, and $E_{1g}$ of point group $D_{6h}$. For a binary combination to be IR-active at the M and K points it must correlate with at least one of the two IR-active symmetries: $A_{2u}$ and $E_{1u}$ of point group $D_{6h}$. **Table 2** lists all the possible binary combinations for the M, K and Γ points and their Raman and IR activity for *2H*-$MoS_2$. Sets of scattering tensors of the Raman-active phonons are also shown [25b]. In order to construct them we reduced the products of binary combinations to their irreducible constituents and utilized the compatibility relations [26, 27] along $\Gamma \xrightarrow{\Sigma} M$, $\Gamma \xrightarrow{\Lambda} K$ and $M \xrightarrow{T} K$ , as is briefly summarized in the supporting information.

All the modes in the M point (also for Γ) are either even or odd with respect to inversion (*g* or *u*, respectively). Hence, as is evidenced from Table 2 the transitions of M point are divided into either Raman or IR activity. For the K point most of the bands show activity on both. Furthermore, most of the M and K points second-order transitions that are active under parallel polarization backscattering configurations ($z(xx)\bar{z}$ , $z(yy)\bar{z}$ in the Porto notation) are also active under perpendicular polarizations ($z(xy)\bar{z}$, $z(yx)\bar{z}$ ). We note that some of the transitions (with $E_{1g}$ symmetry having $\alpha_{zx(xz)}$ and $\alpha_{yz(zy)}$ polarizability tensor components) are expected to be active under off-resonant conditions only under tilted configuration, for which $\boldsymbol{k}\nparallel c$ axis (although we do not exclude their activity under resonant conditions at non-tilted configuration). Hence, besides those transitions that are forbidden as a result of being odd with respect to inversion, all the second-order transitions are Raman allowed. A detailed discussion and analysis of the polarized Raman spectra, in accordance with our extensive analysis, will be given elsewhere.

According to our analysis some of the assignments for second-order transitions at the M point [3] that are currently cited in the literature are not symmetry-allowed. This is because they are combinations of modes with different inversion symmetries (*g* or *u*); below we show the assignment and their symmetry product: $A_{1g}$ (M)+LA(M) ($A_g$x$B_{2u}$), $E^1_{2g}$ ($M_2$)+LA(M) ($A_g$x$B_{2u}$), $E_{1g}$ ($M_1$)+LA(M) and $E_{1g}$ ($M_2$)+LA(M) ($B_{2g}$x$B_{2u}$ and $B_{3g}$x$B_{2u}$, respectively). In contrast, the symmetry products of the combinations that use instead the LA' phonon are symmetry allowed: $A_{1g}$ (M)+LA'(M) ($A_g$x$A_g$), $E^1_{2g}$ ($M_2$)+LA'(M) ($A_g$x$A_g$), $E_{1g}$($M_1$)+LA'(M) and $E_{1g}$ ($M_2$)+LA'(M) ($B_{2g}$x$A_g$ and $B_{3g}$x$A_g$, respectively). However, we bear in mind that an alternative symmetry-allowed ($B_{2u}$x$B_{2u}$) combination with the LA(M) phonon ($E_{1u}$ ($M_1$)+LA(M)) can be found at about similar frequency to that of the symmetry-forbidden combination of $E^1_{2g}$ ($M_2$)+LA(M). The same can also apply to some of the other combinations/difference bands (with $B_{1u}$ (M) phonons) that may potentially contribute at about similar frequencies to those of the $A_{1g}$ (M) phonons. In the resonant spectrum, which will be shown below and in our analysis henceforward, we consider the even-symmetry (with respect to inversion) second-order combinations/difference bands, which involve $A_{1g}$ (M) and $E^1_{2g}$ ($M_2$) (and not $B_{1u}$ (M) and $E^1_{1u}$ ($M_1$)), the dominant ones. Our approach may be substantiated in further studies.

In accordance with the results of Tables 1 and 2, in **Tables 3** and **4** we present a complete set of frequencies of the *2H*-$MoS_2$ second-order Raman- and IR-active combinations, from phonons at M and K at low temperatures (Table 3) and for Γ at 300 K (Table 4). For the Raman-active bands the upper and lower numbers denote combination and difference bands, respectively. The components for different active configurations (while maintaining the respective background colors which denote the various symmetries of Table 3) are also shown.

As mentioned above, unlike the case of the M point, most of the second-order combinations at the K point are both IR- and Raman-allowed. Additionally, the second-order bands that are constructed from *both* phonons of $A_{1g}$ (M), $E^1_{2g}$ ($M_2$), $E^2_{2g}$ ($M_1$) (TA' (M)) and $E^2_{2g}$ ($M_2$) (LA' (M)), which will be referred henceforward as the 'resonant group', are denoted in Table 3 with a thick blue frame (see supporting information for a comment on the nature of $E^1_{2g}$ ($M_1$)). All those combinations are also characterized by temperature-dependent intensity (not shown), clearly indicating their resonant nature (with the A exciton). The focus on those bands allows us to narrow down the possibilities in analyzing the intricate resonant Raman spectrum, which contains about 80 transitions in the spectral range of 80-1130 $cm^{-1}$.

**B. Resonant and off-resonant Raman scattering of bulk *2H*-$MoS_2$**

Shown in **Figure 1** is the Raman spectrum of *2H*-$MoS_2$ at 95 K, measured using excitation energy of 1.96 eV in the spectral range of 80-850 $cm^{-1}$, and divided for clarity into four spectral sub-ranges (a-d). We also show positions of the second-order transitions' bands (active under the backscattering configuration with $\boldsymbol{k} \parallel c$ axis) of origin from M, K and Γ points, (Tables 3 and 4). For Γ, the estimated low-temperature frequencies were obtained from measured values at room temperatures [1-3] after adding of ~ 2.5 $cm^{-1}$ (taking an estimated typical shift of ~ 0.012 $cm^{-1}$/K [5] for first-order phonons). For comparison the room temperature Raman spectrum using excitation energy of 1.58 eV is also shown (after being also shifted in frequencies towards those of lower temperatures for the sake of clarity). In order to highlight the second-order transitions at the M point that are due to the 'resonant group' the respective transitions are denoted by thicker blue marks. The rest of the M point group are denoted in red marks. It is clear from Figure 1 that the majority of the prominent bands are attributed to the 'resonant group' at M. However, some of the prominent bands cannot be attributed to this group and the contribution of 'off- resonant' M and from K and Γ-point phonons must be taken into account. An example is the band at ~ 756 $cm^{-1}$. Shown in **Figure 2** is the off-resonant Raman scattering spectrum around this band, measured at 300 K, and $E_i$ = 1.58 eV. In order to facilitate the discrimination of the large number of potential spectral contributions we compare the Stokes and anti-Stokes spectra (frequency is shown in absolute scale). The calculated frequencies of second-order transitions at M, K and Γ are also shown with the denoted respective colors. Black arrows are pointing to the central frequencies of the contributing bands. Significantly, some of the bands are masked in the Stokes spectrum and are 'exposed' in the corresponding anti-Stokes spectrum. The full spectral profile of the ~756 $cm^{-1}$ band *cannot* be correlated exclusively with the M-point phonons. Clearly, some participation of K phonons and possibly Γ phonons is evident.

Finally, we emphasize that bands, which are very weak under resonant conditions, may become prominent in the spectrum, measured under off-resonant conditions. For example, the $E_{1g}$ $(M_1)$+$E_{1g}$ $(M_2)$ combination may be responsible for the band which appears at ~ 632 $cm^{-1}$ in the low side of the $A_{1g}$ (M)+LA' (M) [9]. This band becomes visible as we move away from resonance with the *A* exciton, either by altering the excitation energy (on both sides of the ~ 1.9 eV resonance), or by increasing the temperature (as will be reported elsewhere).

Shown in **Figure 3** are sequences of all possible contributions (combination and difference bands) rendered in the form of 'flowers' for all the 2nd (circles) and 3rd (squares) order resonant transitions from the four 'resonant group' phonons. Processes that lead to negative shifts (like LA'(M) -$A_{1g}$ (M) ) are taken in their absolute values. At positions where neither is shown there is a process (e.g. $A_{1g}$ (M) +$E^2_{2g}$ $(M_1)$ - $A_{1g}$ (M)), which leads to first-order BZ-edge phonon scattering, and therefor not valid. Blue represents bands that are detected and red denotes bands not detected (possibly due to the low cross sections). Green represents a band that is possibly distinguishable or a band not found, but we believe may possibly be observed under adequate conditions. For example the $A_{1g}$(M) - $E^1_{2g}$ $(M_2)$ that is expected at 42 $cm^{-1}$, below the filter cut-off. Furthermore, we expect this band to be strong, and examination of bulk spectra measured with $E_i$ = 1.96 eV [28] reveals that it may be distinguishable. In fact, we anticipate that under resonance the following bands (and others that are not specified) may appear: 50 $cm^{-1}$ ($E^1_{2g}$ $(M_2)$-2TA' (M)), and 54 $cm^{-1}$ (2LA'(M)-$A_{1g}$ (M)). Some of the bands, marked in green, may be masked by other intense multiphonon bands.

The dominance of the 'resonant group' of phonons is expected particularly in the higher part of the spectrum where higher order transitions may be related exclusively with this group. Presented in **Fig. 4a** is a full set of up to 4th order contributions in the range of 830 -1130 $cm^{-1}$ for $E_i$ = 1.96 eV at 95 K. The denoted bands are constructed from up to 3rd order combinations of $E^1_{2g}$ $(M_2)$ ($\phi_2$) and $A_{1g}$ (M) ($\chi$) phonons, subtracted or added to TA'(M) ($\varphi_1$) or LA'(M) ($\varphi_2$) phonons. In **Figure 4b** the Raman spectra at 300 K for $E_i$ = 1.96 eV and $E_i$ = 1.58 eV are compared. The intensities of the high-order bands decrease with an increase in temperature for $E_i$ = 1.96 eV due to the departure from resonance [5], an observation that also affects lower-order multiphonon intensities [6]. In contrast, for $E_i$ = 1.58 eV we find no spectral features beyond 825 $cm^{-1}$, which is consistent with the attribution of that spectral range to resonant multiphonon processes. Other bands (like $2B^1_{2g}$ (Γ) and $2A_{2u}(\Gamma)$), which could potentially appear in this spectral range (see Table 4), evidently have null (or very minor) contribution.

It is clear from Fig. 3 that, excluding $A_{1g}$ (M)-$E^1_{2g}$ ($M_2$), all the possible second-order combinations of single degenerate phonons at M with an origin at Γ of $A_{1g}$ (Γ), $E^1_{2g}$ (Γ) and $E^2_{2g}$ (Γ) phonons: TA'($\varphi_1$), LA'($\varphi_2$), $E^1_{2g}$ ($M_2$)($\phi_2$), $A_{1g}$($\chi$) are observed in the resonant Raman spectra. In fact, many of the various third-order combinations are detected as well. Since the higher the scattering order the more quantitatively sensitive the spectral analysis, we further checked (beyond the presented spectrum in Fig. 4a) the consistency of the analysis by exploring the correlation between 'expected' and experimental band frequencies for a series of $nA+mB$ ($n$=0-3, $A$=$\chi$, $\phi_2$, $m$=-1,0,1,2,3, $B$=$\varphi_1$, $\varphi_2$). In **Figure 5** we plot $n$A + $m$B *vs.* $m$B for various $n$. We compare the calculated ('expected') and measured frequencies. The correlation between the two is excellent, as is evident from the fact that most of the of 30 bands that are 'expected' (up to the 4$^{th}$ order) within the range of 100 - 1150 cm$^{-1}$ are detected with small discrepancy of ≤ 3 cm$^{-1}$ between the calculated and measured values. Furthermore, higher order transitions are also detected in the room temperature spectrum, which was measured up to ~1400 cm$^{-1}$ (see Fig. S1 presented in the supporting information).

In **Table 5** we list our proposed full assignments (most of which are resonant) of the up to 5$^{th}$ order multiphonon spectra (~80 transitions) measured up to ~1130 cm$^{-1}$ at 95 K. It is important to note that additional bands, many of which are 'masked' by the intense resonant bands, are distinguishable under off-resonant conditions and warrant separate treatment guided by Tables 1-4, that were provided above. Assigning the full spectral range is a complicated task, due to the significant overlap between many of the multiphonon bands that may originate from the Γ, M and K points and to the need to take into account higher-order resonant transitions together with lower-order transitions that might be found in the same spectral position. The ability to be aided by temperature effects is, in some cases limited, under resonant conditions.

In Table 5 we firstly assign the 2$^{nd}$ order M-point bands of the above four phonons and of the ZA' (ι) phonon (see supporting information for a comment on the $E^1_{2g}$($M_1$)($\phi_1$) phonon). The deviation between the measured frequencies and the ones shown in Table 3 is ≤ 3cm$^{-1}$. Then we assign the resonant 3$^{rd}$ order process from the four phonons together with possible other 2$^{nd}$ order M-point processes and also 2$^{nd}$ order K and Γ–points phonons in accordance with Tables 1, 3 and 4. Finally, we assign the spectral range above 825 cm$^{-1}$ (only observed under resonant conditions) with higher-order combinations constructed from M-point 'resonant group' phonons.

An important physical insight from this study, which presents a unified scenario consisting of symmetry analysis and quantitative band frequencies, is that the majority of resonant multiphonon combination processes is between M-point phonons that are from a branch that is optical at Γ. Furthermore, it seems that the cross sections for multiphonon bands that consist of $A_{1g}$ (M) contribution tend to be higher (as can be deduced by the larger portion of detected bands of higher than second order). Although their existence is clear, the dominance of the K contributed transitions is low with respect to those from the M point

## C. The nature of the Raman scattering '2LA band'

From the various bands the one denoted '2LA band' is particularly interesting and following our analysis, this band now calls for reassignment. Previous resonance Raman studies on crystalline $MoS_2$ assigned this band to a second-order process involving the LA(M) phonon. The asymmetry of this peak was assigned to the inverse parabolic shape of the LA(M) dispersion curve near the M point in the BZ [3]. Frey *et al.* [7] suggested that the asymmetric features of the ~ 460 $cm^{-1}$ band is due to a combination of two peaks centered at room temperature at ~454 $cm^{-1}$ (denoted here as $\alpha_1$) and ~465 $cm^{-1}$ ($\alpha_2$). The first is assigned to the 2LA(M) and the second to a Raman-forbidden IR-allowed optical $A_{2u}$ (Γ) mode, which involves asymmetric translation of both Mo and S atoms parallel to the *c* axis [1]. Unlike the $\alpha_2$ band, which can be clearly assigned in Fig. 1, no feasible attribution to the $\alpha_1$ band can be established (see the thick arrow in Fig. 1). Hence, there is a need to explore a different approach in order to trace its origin.

The phonon branches of bulk *2H*-$MoS_2$ in the vicinity of M and K points, calculated by Acata *et al.* [22], are shown in **Figure 6a** with the respective extracted phonon density of states (PDOS). **Figure 6b** compares the spectra around ~460 $cm^{-1}$ for $E_i$ from 1.16 eV to 3.81 eV. In addition, the 2PDOS profile (black line) [22] is also shown (after being shifted, for clarity, by a few $cm^{-1}$). The striking similarity between the ~460 $cm^{-1}$ band of all the spectra taken at $E_i$ other than 1.96 eV and 3.81 eV, and the 2PDOS profile, suggests that the assignment of this band under off-resonant conditions is possibly attributed to combination of BZ edge phonons with additional features found in the PDOS.

In **Figure 6c** the ($\alpha_2/\alpha_1$) intensity ratio is depicted for the set of measurements after adding the respective ratios for 1.16, 1.58, 1.91 [13], 1.96, 2.09 [6] and 2.41 eV. The 2.09 eV measurement is particularly important because it is expected to match exactly the *B* exciton incoming resonance at 300 K. It is evident that in the vicinity of the *A* exciton around ~1.9 eV the ratio increases dramatically. Based on temperature dependent measurements (not shown) which indicate a significantly stronger resonant dependence of $\alpha_2$ than for $\alpha_1$, we conclude that there is a preferred resonance involving the *A* exciton for $\alpha_2$. This, in turn, explains why we see such a strong signal from this band only for ~1.9 eV excitation. The reason for the much stronger excitonic resonance of *A* with respect to that of *B* for the $\alpha_2$ band needs further theoretical clarification.

The resonant nature and characteristics of the high energy side of the of the ~460 $cm^{-1}$ band for the 3.81 eV spectra is fundamentally different from that of the 1.96 eV spectra. For the latter $E_i$, the $\alpha_2$ band is constructed from second-order transitions at the M BZ point with no significant contribution of the $A_{2u}(\Gamma)$ mode. For the former $E_i$ the main contribution, according to a recently published elaborate study of Lee *et al.* [30], comes from the $A_{2u}(\Gamma)$ mode at 470 $cm^{-1}$ (bulk) and $A''_2(\Gamma)$ at ~465 $cm^{-1}$ (monolayer), with the above mentioned 'resonant group' multiphonon contribution being very weak (Fig. S2 of Ref. 30). In fact, it is also shown that the $A_{2u}(\Gamma)$ mode is enhanced for $E_i$=2.81 eV. This may be attributed to a resonant excitation with a conduction band level, presumably positioned in the vicinity of the Γ BZ point [29], which, similar to the intensity of the $A_{2u}(\Gamma)$ band, seems to be particularly sensitive to the number of layers.

In **Figure 7** we show the Lorentzian line fit of the ~ 460 $cm^{-1}$ band in the spectra taken at room temperature with $E_i$ = 1.96 eV (bottom) and 1.58 eV (top). It is evident that this band is comprised of at least five contributions, denoted in the figure as $L_1$-$L_5$ [31]. The $\alpha_1$ band is comprised from $L_1$ and $L_2$ and $\alpha_2$ from $L_3$-$L_5$. The distinction between those contributions is not only enhanced as we shift $E_i$ with respect to the *A* exciton energy, but also as we depart from the resonance upon an increase in temperature. This can be seen in the inset of the lower part of Fig. 7 where we note the high similarity of the '2LA band' spectrum taken at $E_i$ = 1.96 eV and 573 K with the spectrum taken at room temperature and 1.58 eV. Both are off resonant in nature. Other than the red shift of the phonons under high temperatures, the spectral profiles are very similar.

For an appropriate assignment, it is important to realize that the M point is reached at ~ 230 $cm^{-1}$ by two branches. The first is the $E^2_{1u}$ ($M_1$) acoustic branch that reaches the M point with $B_{2u}$ symmetry phonon and the second is the $E^2_{2g}$ ($M_2$) branch that commences at Γ with the $E^2_{2g}$ phonon and reaches M point with $A_g$ symmetry [17]. We denote the former as LA and the latter as LA', with the former being estimated to be slightly (~2 $cm^{-1}$) higher than the latter, according to DFT calculations [23]. The two singly degenerate phonons reach the K point where they maintain their single degeneracy [24]. In line with the correlation with DOS we tentatively assign $L_1$ to a second-order band, possibly $A_{1g}(\Gamma)+E^2_{2g}(\Gamma)$), $L_2$ to a van Hove singularity between K and M (or actually contributions from singularities of two branches that fall at about the same frequency), $L_3$ and $L_4$ to the two phonons (LA' and LA, respectively) at M, and the weak $L_5$ band is tentatively assigned to the weak contributions of the two LA' and LA phonons at K with no significant contribution of $A_{2u}(\Gamma)$. The latter becomes prominent at considerably higher $E_i$ [30] – see Fig 6b for $E_i$=3.81 eV. Finally, the fact that the resonant behavior of the $\alpha_2$ band cannot be exclusively related with 2LA' (M) may suggest that 2LA (M) band is also enhanced near the excitonic resonance. However, it is noteworthy that at lower temperatures (and $E_i$=1.96 eV) the $L_3$ /($L_3$+$L_4$) intensity ratio is somewhat increased (will be shown elsewhere), which may point to the former being slightly more resonant with respect to the latter.

### D. Analysis of infrared-allowed second-order spectrum in bulk *2H*-$MoS_2$

The group theoretical analysis done for the Raman scattering is complemented in Tables 1-4 by the analysis of IR activity. Shown in **Fig. 8** an IR absorption spectrum previously measured by Willson and Yoffe for a 180 μm layer of *2H*-$MoS_2$ (Fig. 39 in [18]). The M point IR-allowed (Raman-forbidden) combinations (Table 3) are also denoted. For the sake of clarity we do not show the K and Γ points IR-active combinations, but we bear in mind that those are also potentially contributors. In black stripes, the absorption bands from Agnihorti [32] are also indicated after subtracting from them 5 $cm^{-1}$ in order to take into account temperature effects (as the reported frequencies [32] are for 77 K). Good correlation between the positions reported in the two studies is evident. We also show proposed assignments to some of the bands. In cases where the M-point phonons of Davydov doublets origins are close in frequencies, there are two allowed combinations with LA/LA' or TA/TA'. For example, $A^1_{2u}$(M) + LA' (M) *vs.* $B^1_{2g}$(M) + LA (M). In the high side of the spectrum there is a band with similar frequency to that found in the Raman spectrum, but with different symmetry (2$A_{1g}$ (M)- Raman active *vs.* $A_{1g}$(M)+$B_{1u}$(M) –IR active).

## E. Symmetry mode analysis of second-order Raman scattering of monolayer *1H*-$MoS_2$

Similar analysis to that of bulk *2H*-$MoS_2$ can be made for monolayer *1H*-$MoS_2$, for which the number of phonon dispersion relations is reduced from 18 to 9. The important differences lie in *i.* The fact that from each of the *2H* Davydov doublets of the optical phonons ($B^1_{2g}$ & $A^1_{2u}$), ($A_{1g}$ & $B_{1u}$), ($E^1_{2g}$ & $E^1_{1u}$) and ($E_{1g}$ & $E_{2u}$) there is one branch left in the Γ point of the *1H* form with $A^{''}_2$, $A^{'}_1$, $E^{'}$ and $E^{''}$ symmetries, respectively, *ii.* The branches at the M point, which commence at Γ and originate from the quasi-acoustic phonons of $E^2_{2g}$ and $B^2_{2g}$ (LA', TA' and ZA' phonons), are absent in the monolayer.

In what follows we show the spectral analysis for the monolayer while restricting ourselves to the Γ and M points. The eigenstates at those points correspond to irreducible representations of the point groups $D_{3h}$ and $C_{2v}$, respectively. **Table 6** lists the various branches with their symmetries at the Γ and M points. In **Table 7** the Raman and IR activity of the various modes are specified and the polarizability tensors of the Raman-active phonons at the Γ point are presented. For the Γ point second-order transitions we find bands, which are either Raman- or IR-active, or both. For the M point all the bands are both Raman- and IR-active. Unlike the phonon energies at Γ, which are mostly known experimentally, at the M point only calculated values are available, which seem to be within a few $cm^{-1}$ from the bulk values [22, 24]. We shall take the corresponding *monolayer* values to be similar to those of the bulk and employ the same procedure described above in tables 1 – 3 (see further details in the supporting information). Although the two are obviously not expected to be truly the same, it may be used as an estimation, in order to provide a guideline for the energies of the second-order transitions of the single layer M and Γ points phonons, which are shown in **Table 8**.

**Figure 9a** presents the Raman spectrum of monolayer $MoS_2$, measured at $E_i$ = 1.96 eV. Apart from the well-known first-order bands [8-10] we focus on the '2LA band' and compare it to the calculated 2PDOS of the monolayer [22] (after being shifted, for clarity, by a few $cm^{-1}$). Two distinct bands are apparent, which are also distinguishable in the 2PDOS. Similar to the bulk case, the correlation between the measured Raman and the 2PDOS is quite good. In line with the correlation with 2PDOS we tentatively assign the lower band to a van Hove singularity between K and M, and the higher one to the 2LA (M) and possibly also to a contribution of 2LA(K) (note the absence of LA' phonons).

Recently, Scheuschner *et al.* [33] showed that the room temperature resonance Raman profile of the $A'_1$ (Γ) phonon fits nicely to the *A* and *B* excitonic transitions for monolayer $MoS_2$. In another recent study Pimenta *et al.* [13] presented the Raman scattering excitation energy dependence for monolayer $MoS_2$, from where it can be shown that the ($\alpha_1/\alpha_2$) intensity ratio follows the *B* excitonic transitions (the lowest excitation energy was 1.95 eV and therefore no conclusion can be drawn for the *A* exciton, which is centered at ~ 1.84 eV). Hence, unlike the bulk, in the monolayer there is an enhancement of the $\alpha_1$ intensity with respect to that of $\alpha_2$. This result cannot be exclusively related with the absence of the LA' phonons in the monolayer because in the bulk $\alpha_2$ does not show as well resonant enhancement in the vicinity of the *B* exciton. Further theoretical clarification is needed to explore the possibility of enhancement in the monolayer of the presumably 2PDOS mode.

In **Fig. 9c** and **9d** we show the spectra of the monolayer in the energy range of and 500-850 $cm^{-1}$ and 110-280 $cm^{-1}$, respectively, where second-order transitions are expected. The former is for overtones and combinations and the latter for the corresponding difference spectra. We denote in the figure the 'expected' positions of the various second-order bands that are due to $A'_1$ (M), E' ($M_1$), E' ($M_2$) and LA (M) modes. By comparing the monolayer spectra with the bulk off-resonant 1.58 eV spectrum, we find that the two have substantial common characteristics and that the measured peaks correlate well with the expected energies listed in Table 8. It is noteworthy that the 228 $cm^{-1}$ band, which is absent in the 1.58 eV bulk spectrum, may be attributed to the $A'_1$ (M)-ZA (M) difference band or, as recently proposed [34], to the LA(M) phonon.

## F. A suggested reassignment of the 'b band' in $MoS_2$

In a resonant Raman scattering study on a single crystal of *2H*-$MoS_2$ at $T = 7$ K and in the laser frequency range of $1.9 < E_i < 2.3$ eV, Sekine *et al.* [4] explored a highly dispersive band at 429 $cm^{-1}$ denoted as the 'b band'. The Stokes peak in that band has been interpreted in terms of a two-phonon process. The first phonon is the $E^1_{1u}$ phonon of finite wavevector and the second a $B^2_{2g}$ quasi-acoustic phonon ($\Delta_2$), which involves vibration of the S-Mo-S planes against each other. According to this interpretation the effect of varying the Inorganic Fullerenes (IF) diameter on the shift of the 'b band' was explained by Frey *et al.* [7], and the effect of pressure and temperature on the 'b band' frequency and the ~5 $cm^{-1}$ shift between the Stokes and anti-Stokes frequencies of this band were analyzed in our previous publication [5].

There are two characteristics that should be fulfilled for the 'b band' interpretation [4] to apply: *i.* The exciting laser line is above the level of the *A* (or *B*) exciton. *ii.* The existence of the $B^2_{2g}$ mode, which appears in the bulk and not in the monolayer. From Fig. 9a it appears that the 'b band' is found in the Raman spectrum of the monolayer. Furthermore, like in the case of the bulk, the center frequency of the 'b band' in the anti-Stokes spectrum also shows a ~ 5 $cm^{-1}$ red shift relative to that found in the Stokes spectrum. Moreover, it seems that the intensity of the 'b band' in the monolayer shows some excitonic resonant dependence (see Fig. 4 in Ref. 13). As a consequence of these new findings (the actual 'appearance' of this band in monolayer and its characteristics), a suggested alternative assignment of the 'b band' is desirable. With the aid of the newly constructed Table 3 we seek for a possible attribution of the transitions that are found at low temperatures in bulk $MoS_2$ at ~ 423 $cm^{-1}$ and ~ 428 $cm^{-1}$ in anti-Stokes and Stokes spectra, respectively [4, 5].

For the M point the only possible attribution is LA (M) + ZA (M) at ~ 417 $cm^{-1}$, which is too low with respect to the experimental values. For the K point we find LA' (K)+TA (K) /TA '(K) at ~ 427 $cm^{-1}$, and LA (K)+TA (K) /TA '(K) at ~ 424 $cm^{-1}$, which are within the expected frequencies and are also shifted by ~ 3 $cm^{-1}$. We also note that LA' (K)+ZA (K)/ZA '(K) and LA (K)+ZA (K)/ZA '(K) (which contain only the $E_{1g}(\Gamma)$ representation in our scheme) are expected at ~ 422 $cm^{-1}$, and ~ 419 $cm^{-1}$, respectively. Although the frequencies at the K point had not been satisfactorily verified to match the experimental data (since most of the attributed transitions are from the M point), we can still suggest that the 'b band' is more likely related with second-order phonon at the K point.

We may therefore suggest that the 'b band' is constructed from two bands, for which their peaks were not well separable in previous studies [4, 5, 7]: the higher band/s (LA' (K)+ TA (K)/TA'(K) or LA (K)+TA (K)/TA '(K)) is resonant with the *A* exciton and is more pronounced in the Stokes outgoing resonance [5], and the lower band is constructed from the contribution/s of combination band/s: LA'(K)+ ZA(K)/ZA'(K) or LA (K)+ZA (K)/ZA '(K). For the monolayer case the phonons LA(K)+TA(K) and LA(K)+ ZA(K) may be considered. Hence, the resonant activation of the upper band (that is minor in the anti-Stokes side) may be related with what we considered as a large blue shift of this mode in the Stokes with respect to the 'less resonant' anti-Stokes side.

Additional support of the proposed reassignment can be found in **Figure 9b,** which compares the Stokes and anti-Stokes spectra of *2H*-$MoS_2$, measured at $E_i$ = 1.58 eV and at room temperature. There are clearly two weak bands at ~ 422 and ~ 417 $cm^{-1}$, in the similar spectral positions of the room temperature 'b band' in the Stokes and anti-Stokes spectra, measured at $E_i$ = 1.96 eV. The appearance of these bands, which are presumably correlated with the 'b band', is not consistent with the full requirements needed for the currently available interpretation [4, 5, 7] to be valid. This is because $E_i$ is significantly lower than the *A* exciton energy. Furthermore, it seems that unlike for the resonant $E_i$=1.96 eV case, the ~ 5 $cm^{-1}$ shift between the two major bands is not observed in the spectra of $E_i$ = 1.58 eV and the two modes appear in about the same spectral positions in both Stokes and anti-Stokes spectra. This new proposed alternative interpretation of the 'b band' remains to be substantiated further. It is particularly important (and challenging) to separate the seemingly hardly-resolved contributions of the sub-bands that construct the ‘b band’ in spectra measured under resonance.

## 4. Conclusions

To summarize, we present a comprehensive analysis of multiphonon Raman spectrum in $MoS_2$. The low temperature resonant spectra were measured with excitation energy of 1.96 eV, which is slightly shifted in energy from the *A* exciton. The analysis consists of symmetry assignments, from which we obtain a broad set of allowed second-order transitions at high-symmetry points in the Brillouin zone.

1. An important physical insight from this study, is that in the bulk the majority of multiphonon resonant bands are proposed to originate from combination processes between four BZ edge phonons at M that are from branches that are optical at Γ (with $A_{1g}$ (Γ), $E^1_{2g}$ (Γ) and $E^2_{2g}$ (Γ)). Consistent with the fact that at the M Brillouin edge only combinations with the same inversion symmetry (*g* or *u*) are Raman-allowed, the contribution of combinations with the LA (M) mode can not be considered with the four phonons that were assigned to construct the ‘resonant group’ $A_{1g}$ (M), $E^1_{2g}$ ($M_2$), $E^2_{2g}$ ($M_1$) (TA’ (M)) and $E^2_{2g}$ ($M_2$) (LA’ (M)). Among those four phonons, all (but one, which is not experimentally detectable in our system) of the $2^{nd}$ order overtones, combination and difference-bands and many of the third order bands are found in the low temperature resonant Raman spectra.

2. As a complemented study we extended the analysis infrared allowed second-order transitions. We also present a multiphonon analysis of the M and Γ points for monolayer *1H*-$MoS_2$. Correlation between the analysis and room temperature Raman spectrum measured at 1.96 eV is satisfactory.

3. We demonstrate that the ‘2LA band’ at ~ 460 $cm^{-1}$ measured at 1.16-2.41 eV is constructed from at least five Lorentzian contributions. Supported by the striking similarity between this band, measured under off- resonant conditions, and the 2PDOS [22], we propose the reassignment of the lower part of the band ($\alpha_1$), that was previously attributed to 2LA(M), to a van Hove singularity between K and M and the higher part ($\alpha_2$) to mostly the overtones of the LA and LA’ phonons at the M point. The $A_{2u}$ (Γ) mode is activated under excitations with considerably higher energy of 3.81 eV (and evidently also with 2.81 eV [30]). Similar approach applies for monolayer *1H*-$MoS_2$.

We anticipate that this analysis will promote the understanding of the currently unresolved mechanism of the multiphonon scattering in $MoS_2$ and its intricate excitation energy dependence. It may as well inform the interpretation of similar processes from a range of other layered dichalcogenides.

## 5. Acknowledgments

We thank Dr. Leila Zeiri from Ben-Gurion University for assistance with acquiring the low temperature and the UV Raman spectra. We also acknowledge the use of the Raman spectrometer under room temperature conditions in the Centralized Research Facilities at Drexel University and Dr. Zhorro Nikolov for his assistance. We gratefully acknowledge Dr. Feng Yan for helping with topographic scanning probe microscopy of the monolayer sample and Vladimir Bačić for sharing his preliminary dispersion curves calculations. Work at Drexel was supported by the National Science Foundation and the Semiconductor Research Corporation (DMR 1124696).

* Corresponding author e-mail: T.Livneh@nrcn.org.il
#Part of the work was done during a Sabbatical leave at the Department of Materials Science & Engineering, Drexel University, USA.

## Tables and Figures

**Table 1** A list of phonons of *2H*-$MoS_2$, their symmetry assignments and frequencies for high-symmetry point of Γ, M (left) and K (right) in the Brillouin zone.

| Band | Γ/$D_{6h}$ | ν($cm^{-1}$)* | Band | | M/$D_{2h}$ | ν($cm^{-1}$)# | Band | | K/$D_{3h}$ | ν($cm^{-1}$)# |
|---|---|---|---|---|---|---|---|---|---|---|
| *A (N)* | $B^1_{2g}$ | 475 | α | $B^1_{2g}$ M | $B_{3g}$ | 393 | a | $B^1_{2g}$ K | A'$_2$ | 380 |
| *B (IR)* | $A^1_{2u}$ | 470 | β | $A^1_{2u}$ M | $B_{1u}$ | 393 | b | $A^1_{2u}$ K | A'$_2$ | 380 |
| *C (R)* | $A_{1g}$ | 409 | χ | $A_{1g}$ M | $A_g$ | 412 | c | $A_{1g}$ K | A'$_1$ | 402 |
| *D (N)* | $B_{1u}$ | 403 | δ | $B_{1u}$ M | $B_{2u}$ | 411 | d | $B_{1u}$ K | A'$_1$ | 402 |
| *E (IR)* | $E^1_{1u}$ | 384 | $ε_1$ | $E^1_{1u}$ $M_1$ | $B_{2u}$ | 370 | e | $E^1_{1u}$ $K_1$ | A'$_1$ | 388 |
| | | | $ε_2$ | $E^1_{1u}$ $M_2$ | $B_{3u}$ | 362 | *g* | $E_{ug}$ $K_2$ | E' | 341 |
| *F (R)* | $E^1_{2g}$ | 383 | $ϕ_1$ | $E^1_{2g}$ $M_1$ | $B_{2u}$ | 362 | | | | |
| | | | $ϕ_2$ | $E^1_{2g}$ $M_2$ | $A_g$ | 370 | f | $E^1_{2g}$ $K_2$ | A'$_1$ | 385 |
| *G (N)* | $E_{2u}$ | 297 | $γ_1$ | $E_{2u}$ $M_1$ | $B_{1u}$ | 338 | h | $E_{2u}$ $K_1$ | A'$_2$ | 338 |
| | | | $γ_2$ | $E_{2u}$ $M_2$ | $A_u$ | 303 | *j* | $E_{ug}$ $K_1$ | E" | 330 |
| *H (R)* | $E_{1g}$ | 286 | $η_1$ | $E_{1g}$ $M_1$ | $B_{2g}$ | 306 | | | | |
| | | | $η_2$ | $E_{1g}$ $M_2$ | $B_{3g}$ | 330 | *i* | $E_{1g}$ $K_2$ | A'$_2$ | 342 |
| *I (N)* | $B^2_{2g}$ | 58 | ι | $B^2_{2g}$ M (ZA') | $B_{3g}$ | 174 | k | $B^2_{2g}$ K (ZA') | A'$_2$ | 184 |
| *J (R)* | $E^2_{2g}$ | 35 | $φ_1$ | $E^2_{2g}$ $M_1$ (TA') | $B_{1g}$ | 160 | l | $E^2_{2g}$ $K_1$ (TA') | A"$_2$ | 188 |
| | | | $φ_2$ | $E^2_{2g}$ $M_2$ (LA') | $A_g$ | 233 | m | $E^2_{2g}$ $K_2$ (LA') | A'$_1$ | 237 |
| *K (AC)* | $E^2_{1u}$ | | $κ_1$ | $E^2_{1u}$ $M_1$ (LA) | $B_{2u}$ | 235 | n | $E^2_{1u}$ $K_2$ (LA) | A'$_1$ | 234 |
| | | | $κ_2$ | $E^2_{1u}$ $M_2$ (TA) | $B_{3u}$ | 156 | o | $E^2_{1u}$ $K_1$ (TA) | A"$_2$ | 190 |
| *L (AC)* | $A^2_{2u}$ | | λ | $A^2_{2u}$ M (ZA) | $B_{1u}$ | 182 | p | $A^2_{2u}$ K (ZA) | A'$_2$ | 185 |

* Measured values at ~300 K [1-3]
# Proposed values at low temperatures (see text)

**Table 2** Group theoretical selection rules for two-phonon Raman and IR activity from the Γ, M and K Brillouin zone points in bulk *2H*-$MoS_2$. The two active groups of symmetries are denoted with different colors. The scattering tensors of the Raman-active phonons are also shown (*c*=*d* away from resonance [25b]).

| Γ | | | | | | M | | | | | | K | | | | | |
|---|---|---|---|---|---|---|---|---|---|---|---|---|---|---|---|---|---|
| Phonon combination | $A_{1g}$ | $E_{1g}$ | $E_{2g}$ | $A_{2u}$ | $E_{1u}$ | Phonon combination | $A_{1g}$ | $E_{1g}$ | $E_{2g}$ | $A_{2u}$ | $E_{1u}$ | Phonon combination | $A_{1g}$ | $E_{1g}$ | $E_{2g}$ | $A_{2u}$ | $E_{1u}$ |
| $A_{1g}xA_{1g}$ | x | | | | | $A_gxA_g$ | x | | x | | | $A'_1xA'_1$ | x | | x | | x |
| $A_{1g}xA_{2u}$ | | | | x | | $A_gxA_u$ | | | | x | | $A'_1xA''_1$ | | x | | | |
| $A_{1g}xB_{2g}$ | | | | | | $A_gxB_{1g}$ | x | | x | | | $A'_1xA'_2$ | | x | | x | |
| $A_{1g}xB_{1u}$ | | | | | | $A_gxB_{1u}$ | | | | x | | $A'_1xA''_2$ | | | x | x | x |
| $A_{1g}xE_{1g}$ | | x | | | | $A_gxB_{2g}$ | | x | | | | $A'_1xE'$ | | | x | | x |
| $A_{1g}xE_{2g}$ | | | x | | | $A_gxB_{2u}$ | | | | | x | $A'_1xE''$ | | x | | | |
| $A_{1g}xE_{1u}$ | | | | | x | $A_gxB_{3g}$ | | x | | | | $A''_1xA''_1$ | x | | x | | x |
| $A_{1g}xE_{2u}$ | | | | | | $A_gxB_{3u}$ | | | | | x | $A''_1xA'_2$ | | | x | x | x |
| $A_{2u}xA_{2u}$ | x | | | | | $A_uxA_u$ | x | | x | | | $A''_1xA''_2$ | | x | | | |
| $A_{2u}xB_{2g}$ | | | | | | $A_uxB_{1g}$ | | | | x | | $A''_1xE'$ | | | x | | x |
| $A_{2u}xB_{1u}$ | | | | | | $A_uxB_{1u}$ | x | | x | | | $A''_1xE''$ | | x | | | |
| $A_{2u}xE_{1g}$ | | | | | x | $A_uxB_{2g}$ | | | | | x | $A'_2xA'_2$ | x | | x | | x |
| $A_{2u}xE_{2g}$ | | | | | | $A_uxB_{2u}$ | | x | | | | $A'_2xA''_2$ | | x | | | |
| $A_{2u}xE_{1u}$ | | x | | | | $A_uxB_{3g}$ | | | | | x | $A'_2xE'$ | | x | | | |
| $A_{2u}xE_{2u}$ | | | x | | | $A_uxB_{3u}$ | | x | | | | $A'_2xE''$ | | | x | | x |
| $B_{2g}xB_{2g}$ | x | | | | | $B_{1g}xB_{1g}$ | x | | x | | | $A''_2xA''_2$ | x | | x | | x |
| $B_{2g}xB_{1u}$ | | | | x | | $B_{1g}xB_{1u}$ | | | | x | | $A''_2xE'$ | | x | | | |
| $B_{2g}xE_{1g}$ | | | x | | | $B_{1g}xB_{2g}$ | | x | | | | $A''_2xE''$ | | | x | | x |
| $B_{2g}xE_{2g}$ | | x | | | | $B_{1g}xB_{2u}$ | | | | | x | $E'xE'$ | x | x | x | | x |
| $B_{2g}xE_{1u}$ | | | | | | $B_{1g}xB_{3g}$ | | x | | | | $E'xE''$ | | x | x | x | x |
| $B_{2g}xE_{2u}$ | | | | | x | $B_{1g}xB_{3u}$ | | | | | x | $E''xE''$ | x | x | x | | x |
| $B_{1u}xB_{1u}$ | x | | | | | $B_{1u}xB_{1u}$ | x | | x | | | | | | | | |

| | (Raman active) | (Raman active) | (Raman active) | (IR active) | (IR active) | | (Raman active) | (Raman active) | (Raman active) | (IR active) | (IR active) |
|---|---|---|---|---|---|---|---|---|---|---|---|
| $B_{1u}xE_{1g}$ | | | | | | $B_{1u}xB_{2g}$ | | | | | x |
| $B_{1u}xE_{2g}$ | | | | | x | $B_{1u}xB_{2u}$ | | x | | | |
| $B_{1u}xE_{1u}$ | | | x | | | $B_{1u}xB_{3g}$ | | | | | x |
| $B_{1u}xE_{2u}$ | | x | | | | $B_{1u}xB_{3u}$ | | x | | | |
| $E_{1g}xE_{1g}$ | x | | x | | | $B_{2g}xB_{2g}$ | x | | x | | |
| $E_{1g}xE_{2g}$ | | x | | | | $B_{2g}xB_{2u}$ | | | | x | |
| $E_{1g}xE_{1u}$ | | | | x | | $B_{2g}xB_{3g}$ | x | | x | | |
| $E_{1g}xE_{2u}$ | | | | | x | $B_{2g}xB_{3u}$ | | | | x | |
| $E_{2g}xE_{2g}$ | x | | x | | | $B_{2u}xB_{2u}$ | x | | x | | |
| $E_{2g}xE_{1u}$ | | | | | x | $B_{2u}xB_{3g}$ | | | | x | |
| $E_{2g}xE_{2u}$ | | | | x | | $B_{2u}xB_{3u}$ | x | | x | | |
| $E_{1u}xE_{1u}$ | x | | x | | | $B_{3g}xB_{3g}$ | x | | x | | |
| $E_{1u}xE_{2u}$ | | x | | | | $B_{3g}xB_{3u}$ | | | | x | |

Raman active

IR active

$$A_{1g} = \begin{bmatrix} a & 0 & 0 \\ 0 & a & 0 \\ 0 & 0 & b \end{bmatrix}$$

$$E_{1g} = \begin{bmatrix} 0 & 0 & -c \\ 0 & 0 & c \\ -d & d & 0 \end{bmatrix}$$

$$E_{2g} = \begin{bmatrix} e & e & 0 \\ e & -e & 0 \\ 0 & 0 & 0 \end{bmatrix}$$

**Table 3** A proposed complete set of second-order phononic transitions from M and K BZ points in *2H*-$MoS_2$. Different groups of Raman scattering tensors are denoted, in accordance with Tables 1 and 2, and are marked with different background colors. The upper number is for a combination and the lower one for a difference band (which are not shown below 70 $cm^{-1}$). In thick blue frames we denote M point Raman allowed resonant second-order processes. The M point IR-allowed combinations are denoted with white background.

| 393 | 174<br>ZA' | 393 | 412 | 411 | 370 | 362 | 362 | 370 | 160<br>TA' | 233<br>LA' | 338 | 303 | 306 | 330 | 235<br>LA | 156<br>TA | 182<br>ZA | | |
|---|---|---|---|---|---|---|---|---|---|---|---|---|---|---|---|---|---|---|---|
| $B^{1}_{2g}$<br>M | $B^{2}_{2g}$<br>M | $A^{1}_{2u}$<br>M | $A_{1g}$<br>M | $B_{1u}$<br>M | $E^{1}_{1u}$<br>$M_1$ | $E^{1}_{1u}$<br>$M_2$ | $E^{1}_{2g}$<br>$M_1$ | $E^{1}_{2g}$<br>$M_2$ | $E^{2}_{2g}$<br>$M_1$ | $E^{2}_{2g}$<br>$M_2$ | $E_{2u}$<br>$M_1$ | $E_{2u}$<br>$M_2$ | $E_{1g}$<br>$M_1$ | $E_{1g}$<br>$M_2$ | $E^{2}_{1u}$<br>$M_1$ | $E^{2}_{1u}$<br>$M_2$ | $A^{2}_{2u}$<br>M | **M**<br>**$D_{2h}$** | |
| 786 | 567<br>219 | 786 | 805 | 804 | 763 | 755 | 755 | 763 | 553<br>233 | 626<br>160 | 731 | 696<br>90 | 699<br>87 | 723 | 628 | 549 | 575 | $B^{1}_{2g}$<br>M | $B_{3g}$ |
| | 348 | 567 | 586<br>238 | 585 | 544 | 536 | 536 | 544<br>196 | 334 | 407 | 512 | 477<br>129 | 480<br>132 | 504<br>156 | 409 | 330 | 356 | $B^{2}_{2g}$<br>M | $B_{3g}$ |
| | | 786 | 805 | 804 | 763 | 755 | 755 | 763 | 553 | 626 | 731 | 696 | 699 | 723 | 628<br>158 | 549<br>237 | 575<br>211 | $A^{1}_{2u}$<br>M | $B_{1u}$ |
| | | | 824 | 823 | 782 | 774 | 774 | 782 | 572<br>252 | 645<br>179 | 750 | 715<br>109 | 718<br>106 | 742<br>82 | 647 | 568 | 594 | $A_{1g}$<br>M | $A_g$ |
| $A^{2}_{2u}$<br>K | 370 | | | 822 | 781 | 773 | 773 | 781 | 571 | 644 | 749<br>73 | 714 | 717 | 741 | 646<br>176 | 567<br>255 | 593<br>229 | $B_{1u}$<br>M | $B_{2u}$ |
| $B^{2}_{2g}$<br>K | 370 | 370 | | | 740 | 732 | 732 | 740 | 530 | 603 | 708 | 673 | 676 | 700 | 605<br>135 | 526<br>212 | 552<br>188 | $E^{1}_{1u}$<br>$M_1$ | $B_{2u}$ |
| $E^{2}_{1u}$<br>$K_1$ | 375 | 375 | 380 | | | 724 | 724 | 732 | 522 | 609 | 700 | 665 | 668 | 692 | 597<br>127 | 518<br>206 | 544<br>180 | $E^{1}_{1u}$<br>$M_2$ | $B_{3u}$ |
| $E^{2}_{2g}$<br>$K_2$ | 375 | 375 | 380 | 380 | | | 724 | 732 | 522 | 595 | 700 | 665 | 668 | 692 | 597<br>127 | 518<br>206 | 544<br>180 | $E^{1}_{2g}$<br>$M_1$ | $B_{2u}$ |
| $E^{2}_{1u}$<br>$K_2$ | 419 | 419 | 424 | 424 | 468 | | | 740 | 530<br>210 | 603<br>137 | 708 | 673 | 676 | 700 | 605 | 526 | 552 | $E^{1}_{2g}$<br>$M_2$ | $A_g$ |
| $E^{2}_{2g}$<br>$K_2$ | 422 | 422 | 427 | 427 | 471 | 474 | | | 320 | 393<br>73 | 498 | 463<br>143 | 466<br>146 | 490<br>170 | 395 | 316 | 342 | $E^{2}_{2g}$<br>$M_1$ | $B_{1g}$ |
| $E_{ug}$<br>$K_1$ | 515<br>145 | 515<br>145 | 520<br>140 | 520<br>140 | 564<br>96 | 567<br>93 | 660 | | | 466 | 571 | 536<br>70 | 539<br>73 | 563<br>97 | 468 | 389 | 415 | $E^{2}_{2g}$<br>$M_2$ | $A_g$ |
| $E_{2u}$<br>$K_1$ | 523<br>153 | 523<br>153 | 528<br>148 | 528<br>148 | 572<br>104 | 575<br>101 | 668 | 676 | | | 676 | 641 | 644 | 668 | 573<br>103 | 494<br>182 | 520<br>156 | $E_{2u}$<br>$M_1$ | $B_{1u}$ |
| $E_{ug}$<br>$K_2$ | 526<br>156 | 526<br>156 | 531<br>151 | 531<br>151 | 575<br>107 | 578<br>104 | 671 | 679 | 682 | | | 606 | 609 | 633 | 538 | 459 | 485 | $E_{2u}$<br>$M_2$ | $B_{3g}$ |
| $E_{1g}$<br>$K_2$ | 527<br>157 | 527<br>157 | 532<br>152 | 532<br>152 | 576<br>108 | 579<br>105 | 672 | 680 | 683 | 684 | | | 612 | 636 | 541 | 462 | 488 | $E_{1g}$<br>$M_1$ | $B_{2g}$ |
| $A^{1}_{2u}$<br>K | 565<br>195 | 565<br>195 | 570<br>190 | 570<br>190 | 614 | 617 | 710 | 718 | 721 | 722 | 760 | | | 660 | 565 | 486 | 512 | $E_{1g}$<br>$M_2$ | $B_{3g}$ |
| $B^{1}_{2g}$<br>K | 565<br>195 | 565<br>195 | 570<br>190 | 570<br>190 | 614<br>146 | 617<br>143 | 710 | 718 | 721 | 722 | 760 | 760 | | | 470 | 389<br>79 | 417 | $E^{2}_{1u}$<br>$M_1$ | $B_{2u}$ |
| $E^{1}_{2g}$<br>$K_2$ | 570<br>200 | 570<br>200 | 575<br>195 | 575<br>195 | 619<br>151 | 622<br>148 | 715 | 723 | 726 | 727 | 765 | 765 | 770 | | | 312 | 338 | $E^{2}_{1u}$<br>$M_2$ | $B_{3u}$ |
| $E^{1}_{1u}$<br>$K_1$ | 573<br>203 | 573<br>203 | 578<br>198 | 578<br>198 | 622<br>154 | 625<br>151 | 718 | 726 | 729 | 730 | 768 | 768 | 773 | 776 | | | 364 | $A^{2}_{2u}$<br>M | $B_{1u}$ |
| $B_{1u}$<br>K | 587<br>217 | 587<br>217 | 592<br>212 | 592<br>212 | 633<br>168 | 636<br>165 | 732<br>72 | 740 | 743 | 744 | 782 | 782 | 787 | 791 | 804 | | | | |
| $A_{1g}$<br>K | 587<br>217 | 587<br>217 | 592<br>212 | 592<br>212 | 633<br>168 | 639<br>165 | 732<br>72 | 740 | 743 | 744 | 782 | 782 | 787 | 791 | 804 | 804 | | | |
| **K**<br>**$D_{3h}$** | $A^{2}_{2u}$<br>K | $B^{2}_{2g}$<br>K | $E^{2}_{1u}$<br>$K_1$ | $E^{2}_{2g}$<br>$K_2$ | $E^{2}_{1u}$<br>$K_2$ | $E^{2}_{2g}$<br>$K_2$ | $E_{ug}$<br>$K_1$ | $E_{2u}$<br>$K_1$ | $E_{ug}$<br>$K_2$ | $E_{1g}$<br>$K_2$ | $A^{1}_{2u}$<br>K | $B^{1}_{2g}$<br>K | $E^{1}_{2g}$<br>$K_2$ | $E^{1}_{1u}$<br>$K_1$ | $B_{1u}$<br>K | $A_{1g}$<br>K | | | |
| | 185<br>ZA | 185<br>ZA' | 190<br>TA | 190<br>TA' | 234<br>LA | 237<br>LA' | 330 | 338 | 341 | 342 | 380 | 380 | 385 | 388 | 402 | 402 | | | |
| | $A_2$' | $A_2$' | $A_2$'' | $A_2$'' | $A_1$' | $A_1$' | E'' | $A_2$' | E' | $A_2$' | $A_2$' | $A_2$' | $A_1$' | $A_1$' | $A_1$' | $A_1$' | | | |

| |
|---|
| **$E_{1g}$** |
| **$E_{2g}$** |
| **$A_{1g}$, $E_{2g}$** |
| **$E_{1g}$,, $E_{2g}$** |
| **$A_{1g}$, $E_{1g}$, $E_{2g}$** |

**Table 4** Raman (colored) and IR (white) active combinations for second-order processes from phonons at Γ in *2H*-$MoS_2$ measured at 300 K. Different groups of Raman scattering tensors are denoted, in accordance with Tables 1 and 2 and marked with different background colors. The upper number is for a combination and the lower one for a difference band (which are not shown below 70 $cm^{-1}$). The x sign denotes an inactive combination.

| $A_{1g}$ 409 | $A_{2u}$ 470 | $B^1_{2g}$ 475 | $B^2_{2g}$ 58 | $B_{1u}$ 403 | $E^1_{2g}$ 383 | $E^2_{2g}$ 35 | $E_{1g}$ 286 | $E_{1u}$ 384 | $E_{2u}$ 297 | **Γ/$D_{6h}$** |
|---|---|---|---|---|---|---|---|---|---|---|
| 818 | 879 | x | x | x | 792 | 444<br>374 | 695<br>123 | 793 | x | $A_{1g}$ 409 |
| | 940 | x | x | x | x | x | 756 | 854<br>86 | 767<br>173 | $A_{2u}$ 470 |
| | | 950 | 533<br>417 | 878 | 858<br>92 | 510<br>440 | 761<br>189 | x | 772 | $B^1_{2g}$ 475 |
| | | | 116 | 461 | 441<br>325 | 93 | 344<br>228 | x | 355 | $B^2_{2g}$ 58 |
| | | | | 806 | 786 | 438 | x | 787 | 800<br>106 | $B_{1u}$ 403 |
| | | | | | 766 | 418<br>348 | 669<br>97 | 767 | 680 | $E^1_{2g}$ 383 |
| | $A_{1g}$ | | | | | 70 | 321<br>251 | 419 | 332 | $E^2_{2g}$ 35 |
| | $A_{1g}$, $E_{2g}$ | | | | | | 572 | 670 | 583 | $E_{1g}$ 286 |
| | $E_{1g}$ | | | | | | | 768 | 681<br>87 | $E_{1u}$ 384 |
| | $E_{2g}$ | | | | | | | | 594 | $E_{2u}$ 297 |

**Table 5** Proposed assignments of *2H*-$MoS_2$ at low temperature (95 K), measured up to 1130 $cm^{-1}$ using excitation energy of 1.96 eV. The notations are shown in Table 1*.

| ν($cm^{-1}$) | Assignment | ν($cm^{-1}$) | Assignment | ν($cm^{-1}$) | Assignment |
|---|---|---|---|---|---|
| 88 | $\varphi_1{}^2\overline{\varphi_2}$ | 405.5 | $D$ | 739 | $\phi_2{}^2$ |
| 92 | $\overline{\varphi_1}^2\chi$ | 411 | $C$ | 756 | $\beta\epsilon_2$ |
| 96 | $\varphi_2{}^2\overline{\phi_2}$ | 428 | # | 768 | $f^2$ |
| 115 | $I^2$ | 456 | $ | 781.5 | $\phi_2\chi$ |
| 118 | $\varphi_1\phi_2\overline{\chi}$ | 466 | $\varphi_2{}^2$ | 788 | $\varphi_1{}^2\varphi_2{}^2, \alpha^2, \beta^2$ |
| 125 | $\gamma_2\overline{\iota}, \varepsilon_2\overline{\kappa_1}$ | 470 | $\kappa_1{}^2$ | 794.5 | de, $ce$ |
| 137 | $\overline{\varphi_2}\phi_2$ | 478 | $k^2$ | 802 | $\chi\alpha$ |
| 142 | $\varphi_1{}^2\varphi_2\overline{\chi}$ | 479 | $\varphi_1{}^3$ | 808 | $\varphi_1\varphi_2\chi$ |
| 151 | $h\overline{I}$ | 485 | $\overline{\varphi_1}\varphi_2\chi$ | 824 | $\chi^2, C^2, \delta^2$ |
| 159 | $\gamma_1\overline{\lambda}$ | 501.5 | $\overline{\varphi_1}^2\chi^2$ | 836 | $\varphi_2{}^2\phi_2$ |
| 168 | $c\overline{m}$ | 514.5 | - | 850 | $\varphi_1{}^3\phi_2$ |
| 180.5 | $\varphi_2\overline{\chi}$ | 530 | $\varphi_1\phi_2$ | 864 | $\overline{\phi_2}\chi^3$ |
| 192.5 | $\phi_2\overline{\iota}$ | 546 | $\phi_2\iota$ | 877 | $\varphi_2{}^2\chi$ |
| 202 | $\varphi_1\overline{\phi_2}\chi$ | 560.5 | $\varphi_2\phi_2{}^2\overline{\chi}$ | 892 | $\varphi_1{}^3\chi$ |
| 210 | $\overline{\varphi_1}\phi_2$ | 566 | $\iota\alpha$ | 901-947 | $\varphi_2{}^4, \varphi_1\phi_2\chi$ .... |
| 229 | $\delta\overline{\lambda}, H\overline{I}$ | 573 | $\varphi_1\chi$ | 958 | $\overline{\varphi_2}\phi_2\chi^2$ |
| 236.5 | $\overline{\iota}\chi$ | 576 | $\beta\lambda, H^2$ | 973 | $\varphi_2\phi_2{}^2$ |
| 245 | $\delta\overline{\kappa_2}$ | 581 | $\overline{\varphi_1}\phi_2{}^2$ | 984 | $\varphi_1\chi^2$ |
| 254 | $\overline{\varphi_1}\chi$ | 591 | $\overline{\varphi_2}\chi^2$ | 1000 | $\overline{\varphi_2}\chi^3$ |
| 312 | $\kappa_2{}^2$ | 603 | $\varphi_2\phi_2$ | 1013 | $\varphi_2\phi_2\chi$ |
| 319.5 | $\varphi_1{}^2$ | 620 | $\overline{\varphi_1}\phi_2\chi$ | 1026 | $\varphi_2{}^3\phi_2{}^2\overline{\chi}$ |
| 329 | $\phi_2{}^2\overline{\chi}$ | 626 | $\varphi_1\varphi_2{}^2, \varphi_2\alpha$ | 1030 | $\overline{\varphi_1}\phi_2\chi^2$ |
| 340 | $\varphi_1\overline{\varphi_2}\chi, \kappa_2\lambda$ | 632 | $cm, \eta_1\eta_2$ | 1043 | $\overline{\varphi_1}\varphi_2{}^2\phi_2{}^2$ |
| 347 | $\iota^2$ | 646 | $\varphi_2\chi$ | 1055 | $\varphi_2\chi^2$ |
| 356 | $\overline{\varphi_2}^2\chi^2$ | 662 | $\overline{\varphi_1}\chi^2$ | 1061 | $\varphi_1{}^2\phi_2{}^2$ |
| 365 | $\lambda^2$ | 677 | $\gamma_1{}^2$ | 1071 | $\varphi_2{}^3\phi_2$ |
| 383 | $l^2$ ,$o^2$ ? | 684 | $i^2$ | 1111 | $\varphi_2{}^3\chi, \phi_2{}^3$ |
| 386.5 | $F$ | 691 | $\varphi_1{}^2\phi_2$ | 1117 | - |
| 392.5 | $\kappa_1\kappa_2$ | 700 | $\varphi_2{}^3$ | 1130 | $\varphi_1\varphi_2\phi_2{}^2$ |
| 394.5 | $\varphi_1\varphi_2$ | 709 | - | | |
| 398.5 | - | 725.5 | $\phi_1{}^2$ | | |

* $x$ and $\overline{x}$ stand for addition and subtraction of a first-order *x* phonon and n in $x^n$ represents the order of the multiphonon transition of the *x* phonon.

# The 'b band' [4, 5, 7] – see below in Section 3F.

$ See below the attribution of the '2LA band' in Section 3C.

**Table 6** A list of phonons of *1H*-$MoS_2$, their symmetry assignments and frequencies for Γ and M points in the Brillouin zone.

| Band | $\Gamma/D_{3h}$ | $\nu(cm^{-1})$ | Band | #$\nu(cm^{-1})$ | $M/C_{2v}$ |
|---|---|---|---|---|---|
| *A (IR)* | $A''_2$ | ~470** | $A''_2M$ | 393 | $B_1$ |
| *B (R)* | $A'_1$ | 403* | $A'_1M$ | 412 | $A_1$ |
| *C (IR+R)* | *E'* | 384* | $E'M_1$ | 370 | $A_1$ |
| | | | $E'M_2$ | 362 | $B_2$ |
| *D (R)* | *E''* | ~273** | $E''M_1$ | 306 | $A_2$ |
| | | | $E''M_2$ | 330 | $B_1$ |
| *E (AC)* | *E'* | 0 | $E'M_1$ (LA) | 235 | $A_1$ |
| | | | $E'M_2$ (TA) | 160 | $B_2$ |
| *F (AC)* | $A''_2$ | 0 | $A''_2M$ (ZA) | 182 | $B_1$ |

Measured (*) [8-10] or estimated (**) values at ~300 K.
#The frequencies for the M point are estimated.

**Table 7** Group theoretical selection rules for two-phonon Raman and IR activity from the Γ and M Brillouin zone points of monolayer *1H*-$MoS_2$. The three active groups of symmetries are denoted with different colors. The scattering tensors of the Raman-active phonons are also shown (*c*=*d* away from resonance [25b]).

| **Γ** | | | | | **M** | | | | |
|---|---|---|---|---|---|---|---|---|---|
| Phonon combination | $A'_1$ | $E''$ | $E'$ | $A''_2$ | Phonon combination | $A'_1$ | $E''$ | $E'$ | $A''_2$ |
| $A'_1xA'_1$ | X | | | | $A_1xA_1$ | X | X | X | |
| $A'_1xA''_1$ | | | | | $A_1xA_2$ | | X | X | X |
| $A'_1xA'_2$ | | | | | $A_1xB_1$ | X | X | X | |
| $A'_1xA''_2$ | | | | X | $A_1xB_2$ | | X | X | X |
| $A'_1xE'$ | | | X | | $A_2xA_2$ | X | X | X | |
| $A'_1xE''$ | | X | | | $A_2xB_1$ | | X | X | X |
| $A''_1xA''_1$ | X | | | | $A_2xB_2$ | X | X | X | |
| $A''_1xA'_2$ | | | | X | $B_1xB_1$ | X | X | X | |
| $A''_1xA''_2$ | | | | | $B_1xB_2$ | | X | X | X |
| $A''_1xE'$ | | X | | | $B_2xB_2$ | X | X | X | |
| $A''_1xE''$ | | | X | | | | | | |
| $A'_2xA'_2$ | X | | | | Raman active (blue) | | | | |
| $A'_2xA''_2$ | | | | | Raman & IR active (green) | | | | |
| $A'_2xE'$ | | | X | | IR active (orange) | | | | |
| $A'_2xE''$ | | X | | | | | | | |
| $A''_2xA''_2$ | X | | | | | | | | |
| $A''_2xE'$ | | X | | | | | | | |
| $A''_2xE''$ | | | X | | | | | | |
| $E'xE'$ | X | | X | | | | | | |
| $E'xE''$ | | X | | X | | | | | |
| $E''xE''$ | X | | X | | | | | | |

$$A'_1 = \begin{bmatrix} a & 0 & 0 \\ 0 & a & 0 \\ 0 & 0 & b \end{bmatrix}$$

$$E'' = \begin{bmatrix} 0 & 0 & -c \\ 0 & 0 & c \\ -d & d & 0 \end{bmatrix}$$

$$E' = \begin{bmatrix} e & e & 0 \\ e & -e & 0 \\ 0 & 0 & 0 \end{bmatrix}$$

**Table 8** A complete set of second-order phononic transitions from M and Γ Brillouin zone points in monolayer *1H*-$MoS_2$. The upper number is for a combination and the lower one for a difference band (which are not shown below 70 $cm^{-1}$). *The frequencies of the M point are the estimated ones* (based on the assumed close resemblance of the *1H* and *2H* frequencies). Different groups of Raman scattering tensors are denoted, in accordance with Tables 6 and 7 and are marked with different background colors.

| | | 393 | 412 | 370 | 362 | 306 | 330 | 235 | 160 | 182 | |
|---|---|---|---|---|---|---|---|---|---|---|---|
| | | $A''_2$<br>*M* | $A'_1$<br>*M* | $E'$<br>$M_1$ | $E'$<br>$M_2$ | $E''$<br>$M_1$ | $E''$<br>$M_2$ | $E'$<br>$M_1$ | $E'$<br>$M_2$ | $A''_2$<br>*M* | **M**<br>$\mathbf{C_{2v}}$ |
| | | 786 | 805 | 763 | 755 | 699<br>87 | 723 | 628<br>158 | 553<br>233 | 575<br>211 | $A''_2$<br>*M* |
| | | | 824 | 782 | 774 | 718<br>106 | 742<br>82 | 647<br>177 | 572<br>252 | 594<br>230 | $A'_1$<br>*M* |
| | | | | 740 | 732 | 676 | 700 | 605<br>135 | 530<br>210 | 552<br>188 | $E'$<br>$M_1$ |
| | | | E'', E' | | 724 | 668 | 706 | 597<br>127 | 522<br>202 | 544<br>180 | $E'$<br>$M_2$ |
| | | | | A'1, E'',<br>E' | | 612 | 636 | 541<br>71 | 466<br>146 | 488<br>124 | $E''$<br>$M_1$ |
| 470 | $A''_2$ | 940 | A'1, E' | | | | 660 | 565<br>95 | 490<br>170 | 512<br>148 | $E''$<br>$M_2$ |
| 403 | $A'_1$ | 873 | 806 | E' | | | | 470 | 395<br>75 | 417 | $E'$<br>$M_1$ |
| 384 | $E'$ | 854<br>86 | 787 | 768 | E'' | | | | 320 | 342 | $E'$<br>$M_2$ |
| 280 | $E''$ | 750<br>190 | 783<br>127 | 664<br>104 | 560 | A'1 | | | | 364 | $A''_2$<br>*M* |
| | **Γ**<br>$\mathbf{D_{3h}}$ | $A''_2$ | $A'_1$ | $E'$ | $E''$ | | | | | | |

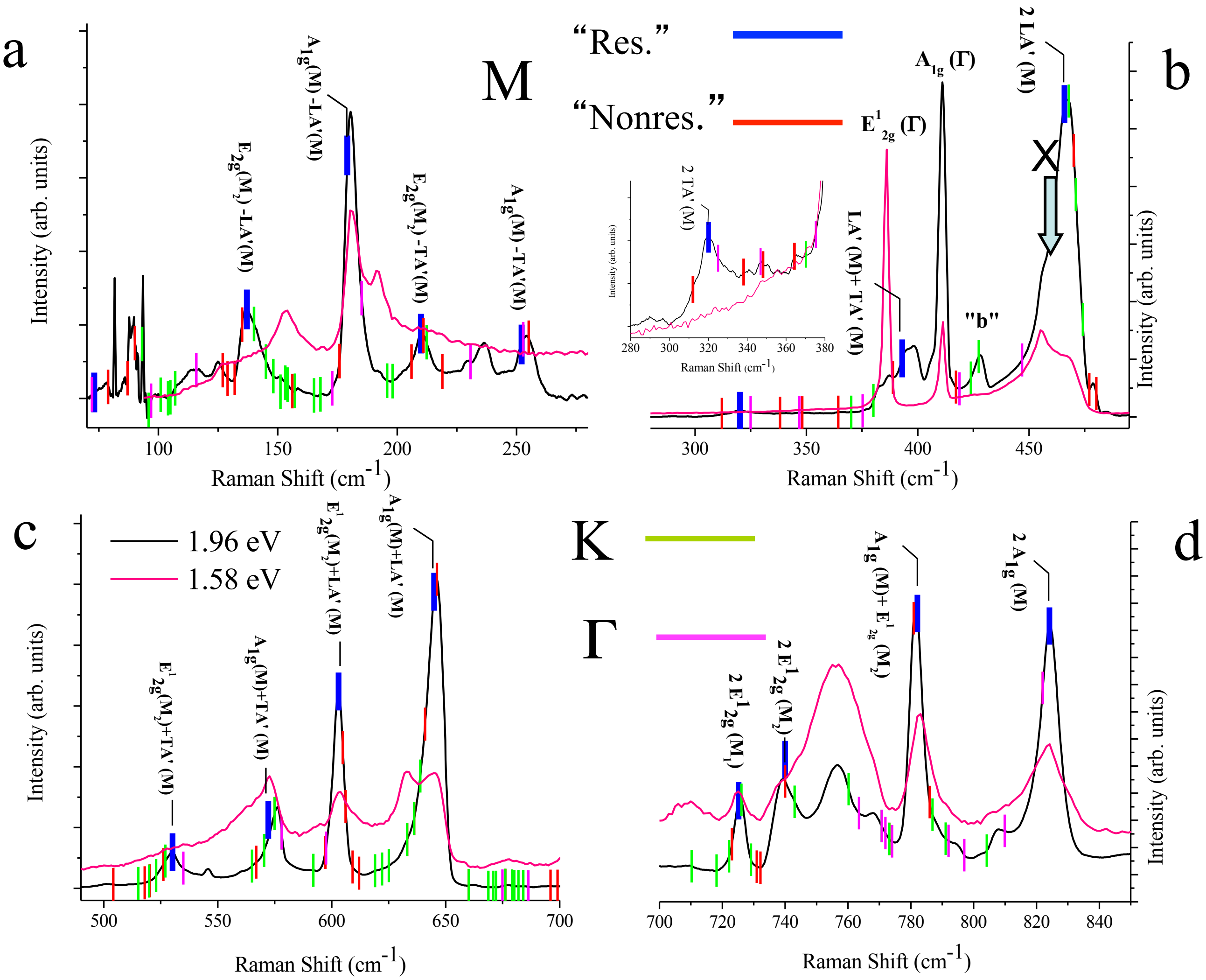

**Figure 1** The resonant Raman spectrum of *2H*-$MoS_2$ at 95 K measured using excitation energy of 1.96 eV, divided for clarity into four spectral sub-ranges (a-d). Positions of the second-order transitions' bands of origin from M ('Res'-blue, 'Nonres'-red), K (green) and Γ (magneta) Brillouin zone points, that are active under back scattering configuration, are also denoted. For comparison the room temperature Raman spectrum using excitation energy of 1.58 eV is also shown (pink). Note that for the band at ~ 456 $cm^{-1}$, there is no possible assignment that may point to its origin.

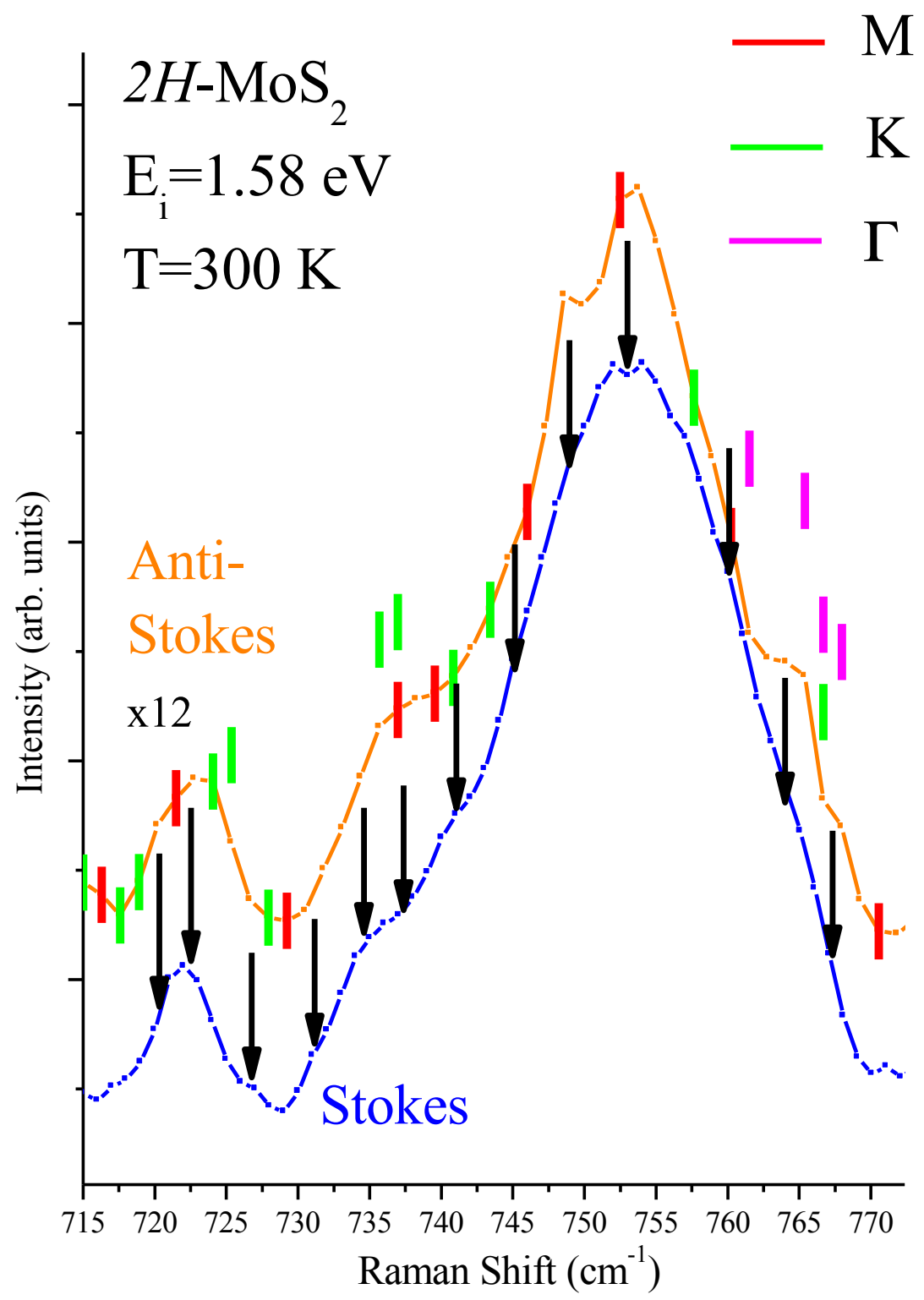

**Figure 2** A focus on the spectral profile around the band at ~756 $cm^{-1}$ measured with $E_i$=1.58 eV and 300 K. In order to facilitate the differentiation of the large number of potential spectral contributions we show Stokes and anti-Stokes spectra (shown in absolute frequency and scaled to similar intensities). Black arrows point to the central frequencies of the contributing bands. The calculated frequencies of second-order transition (including those attributed in Table 3 to have $E_{1g}$ symmetry) at M, K and Γ points are also shown with the denoted respective colors.

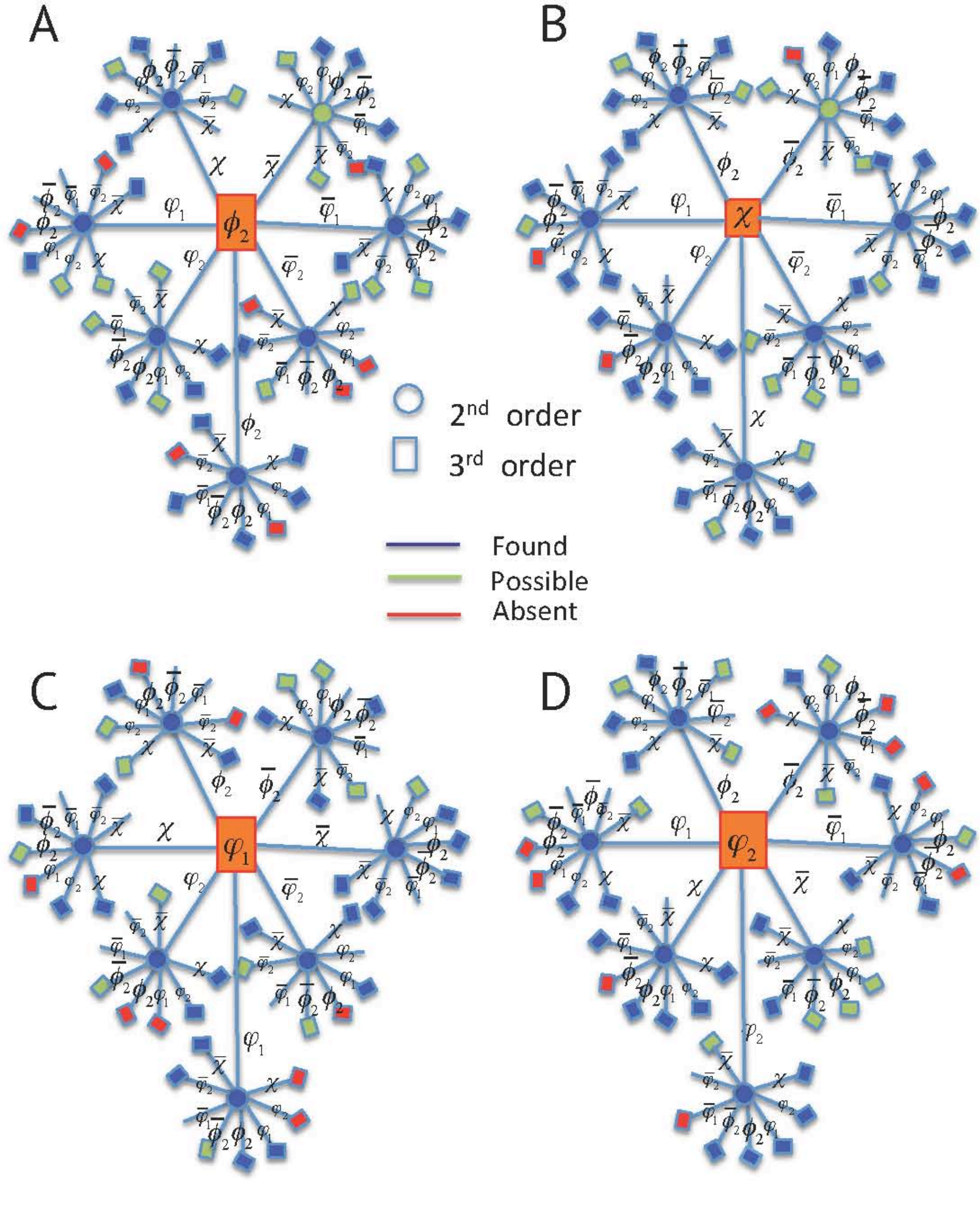

$\chi$ – $A_{1g}$(M), $\phi_2$ – $E^1_{2g}(M_2)$, $\varphi_1$ – TA' (M), $\varphi_2$ – LA' (M)

**Figure 3** Sequences of all possible contributions (combination and difference bands) are shown in the form of 'flowers' for all the 2nd (circles) and 3rd (squares) order resonant transitions from the $E_{2g}$ ($M_2$)($\phi_2$) (A), $A_{1g}$ ($\chi$) (B) and TA'(M) ($\varphi_1$) (C), LA'(M) ($\varphi_2$) (D), phonons. Blue represents bands that are detected and red denotes bands that are not detected. Green represents a band that is possibly distinguishable or a band that although not found, we believe one that may possibly be observed under adequately sensitive experimental conditions.

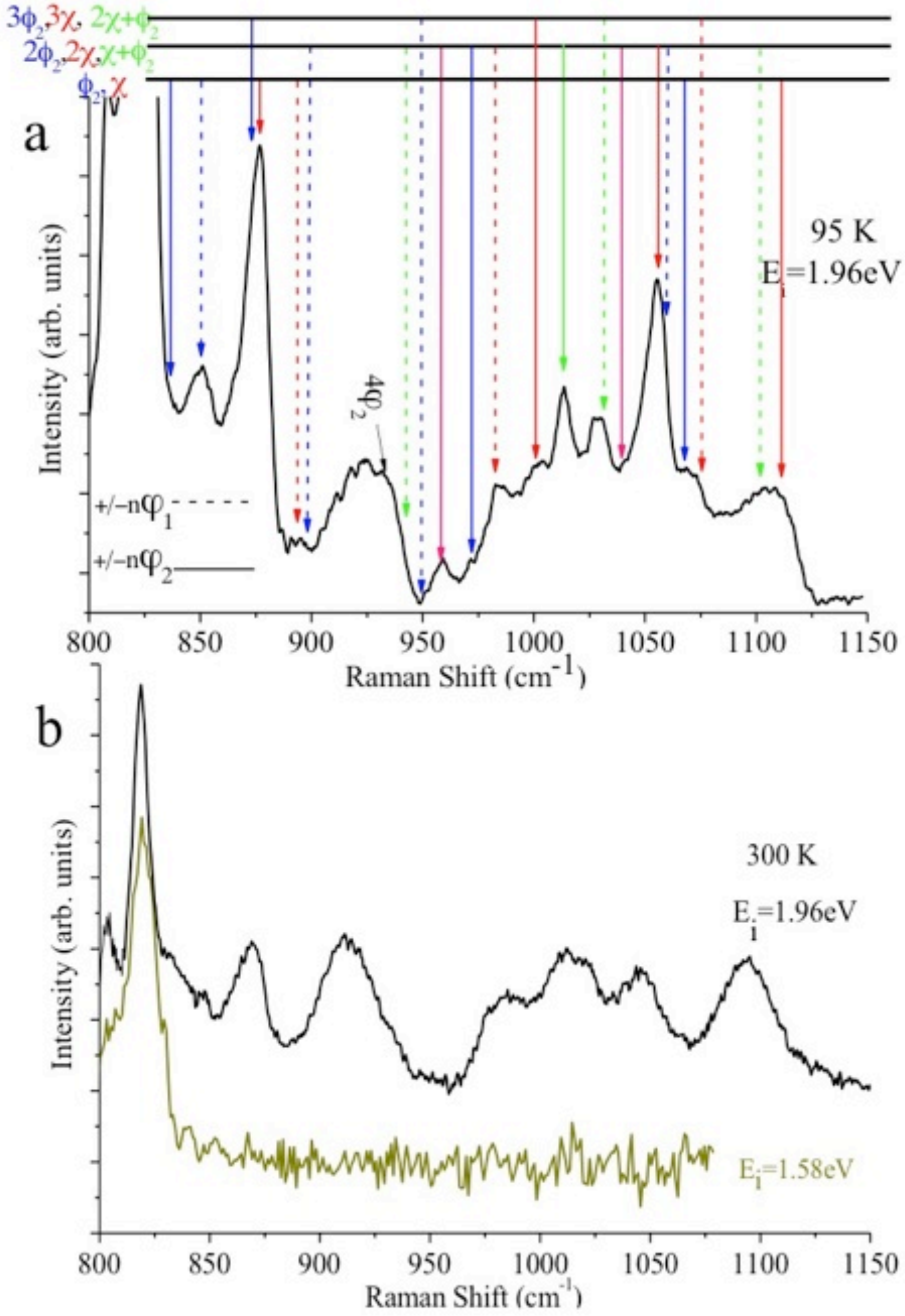

**Figure 4** a) A full set of up to 4th order contributions in the range of 830 -1130 $cm^{-1}$ for $E_i$ = 1.96 eV at 95 K. The bands are constructed from up to 3rd order combinations of $E^1_{2g}$ ($M_2$)($\phi_2$) (blue) and $A_{1g}$ (M) ($\chi$) (red) phonons subtracted or added to TA' (M) ($\varphi_1$) (dashed line) or LA' (M) ($\varphi_2$) (solid line) phonons. Combinations of $E^1_{2g}$ ($M_2$) +$A_{1g}$ (M) are shown in green. b) A comparison of Raman spectra at 300 K for $E_i$ = 1.96 eV and $E_i$ = 1.58 eV.

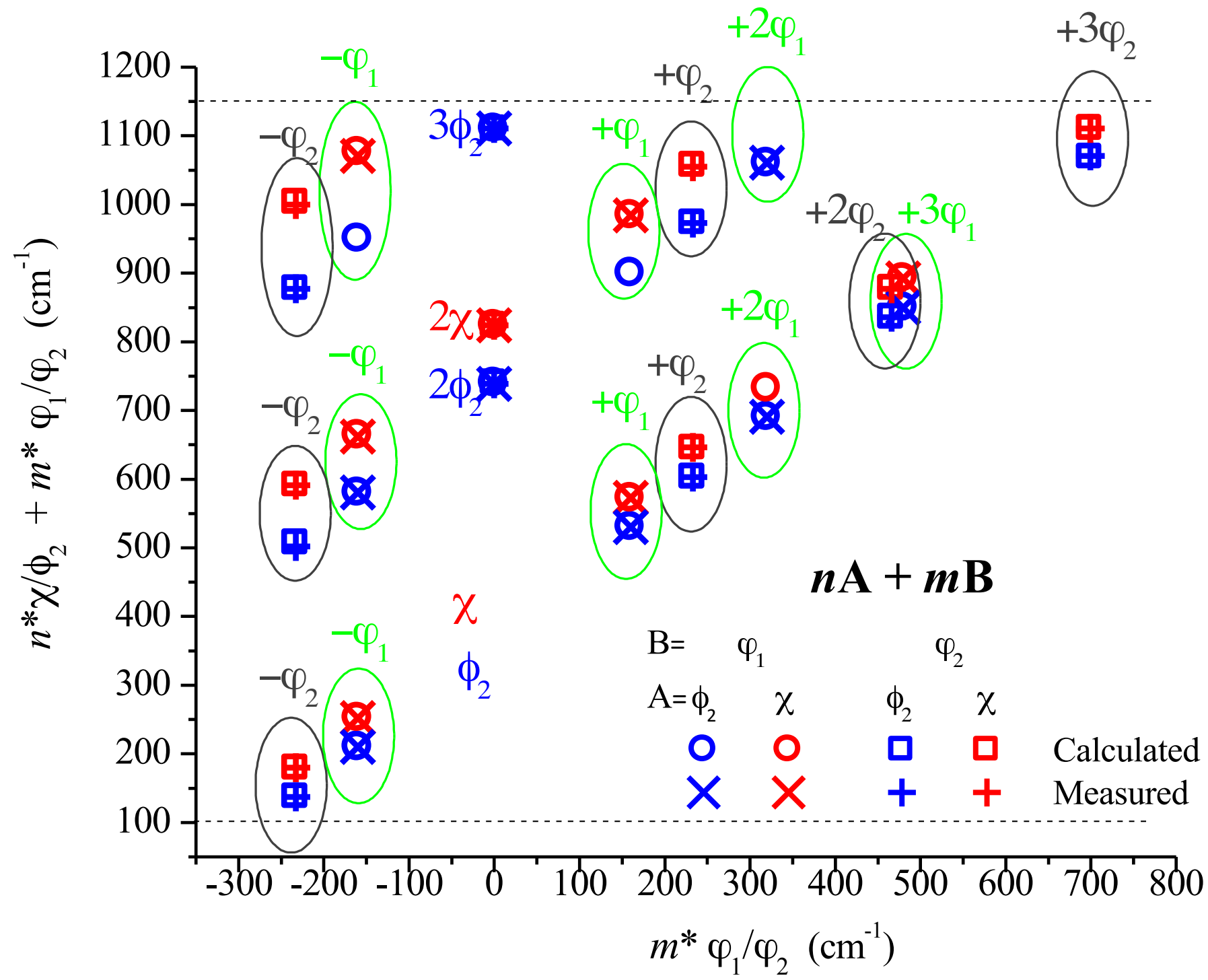

**Figure 5** A plot of *n*A + *m*B *vs.* *m*B ($n$ = 0-3, $A = \chi$, $\phi_2$, $m$ = -1,0,1,2,3, $B = \varphi_1$, $\varphi_2$) for various $n$ of the calculated and the measured frequencies.

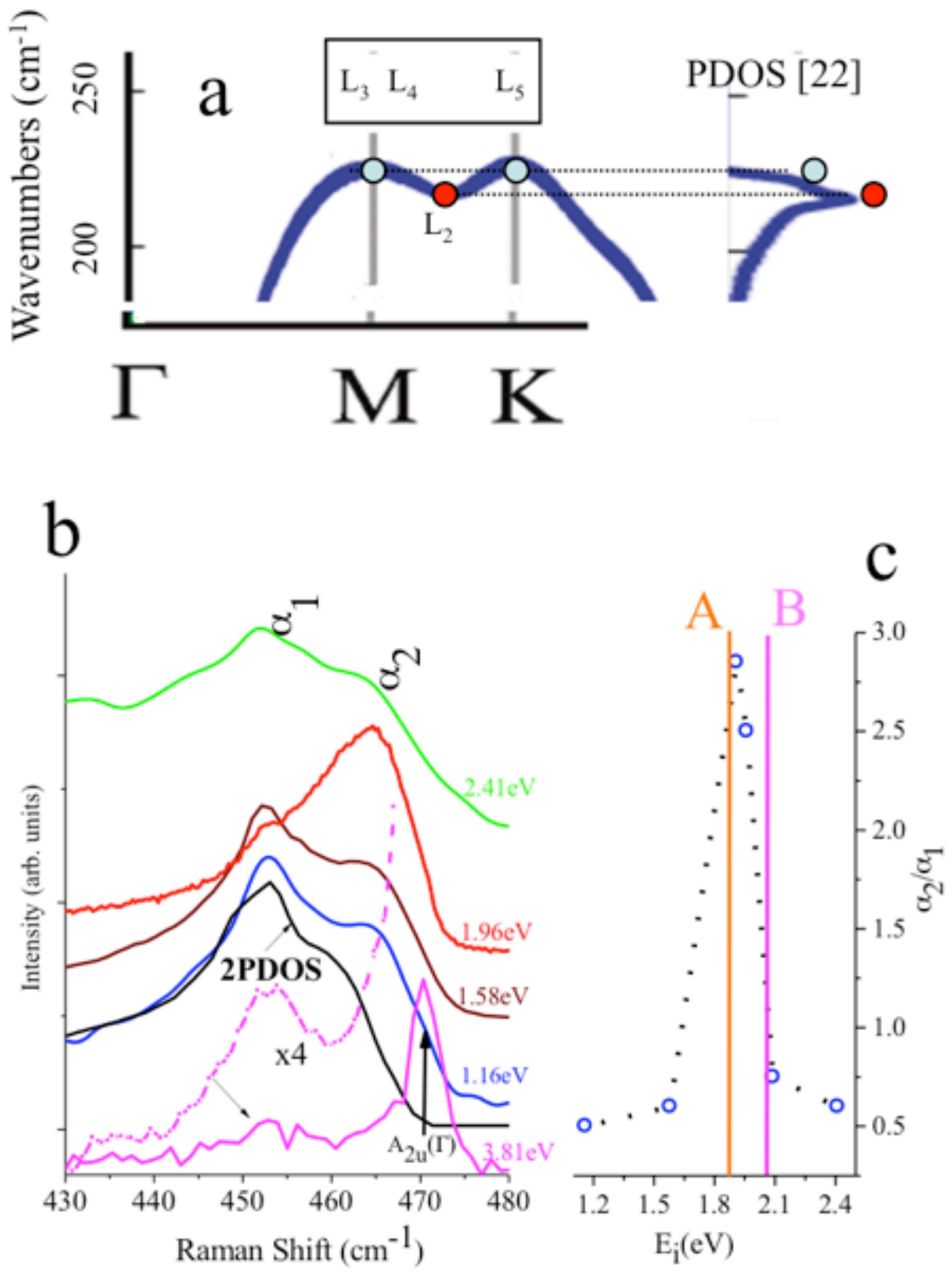

**Figure 6** a) The phonon branches in the vicinity of M and K points, calculated by Acata *et al.* [22], with the respective extracted phonon density of states (PDOS) of *2H*-$MoS_2$. b) A comparison of the spectra around ~ 460 $cm^{-1}$ for various excitation energies $E_i$ from 1.16 eV to 3.81 eV (dashed dotted line - a zoom of the smoothed lower band). The 2PDOS profile [22] is shown in black line. c) The ($\alpha_1/\alpha_2$) intensity ratio is depicted for the set of measurements after adding the respective ratios for 1.16, 1.58, 1.91 (taken from [13]), 1.96, 2.09 eV (taken from [6]) and 2.41 eV. The room temperature energies of the *A* and *B* excitons are also denoted.

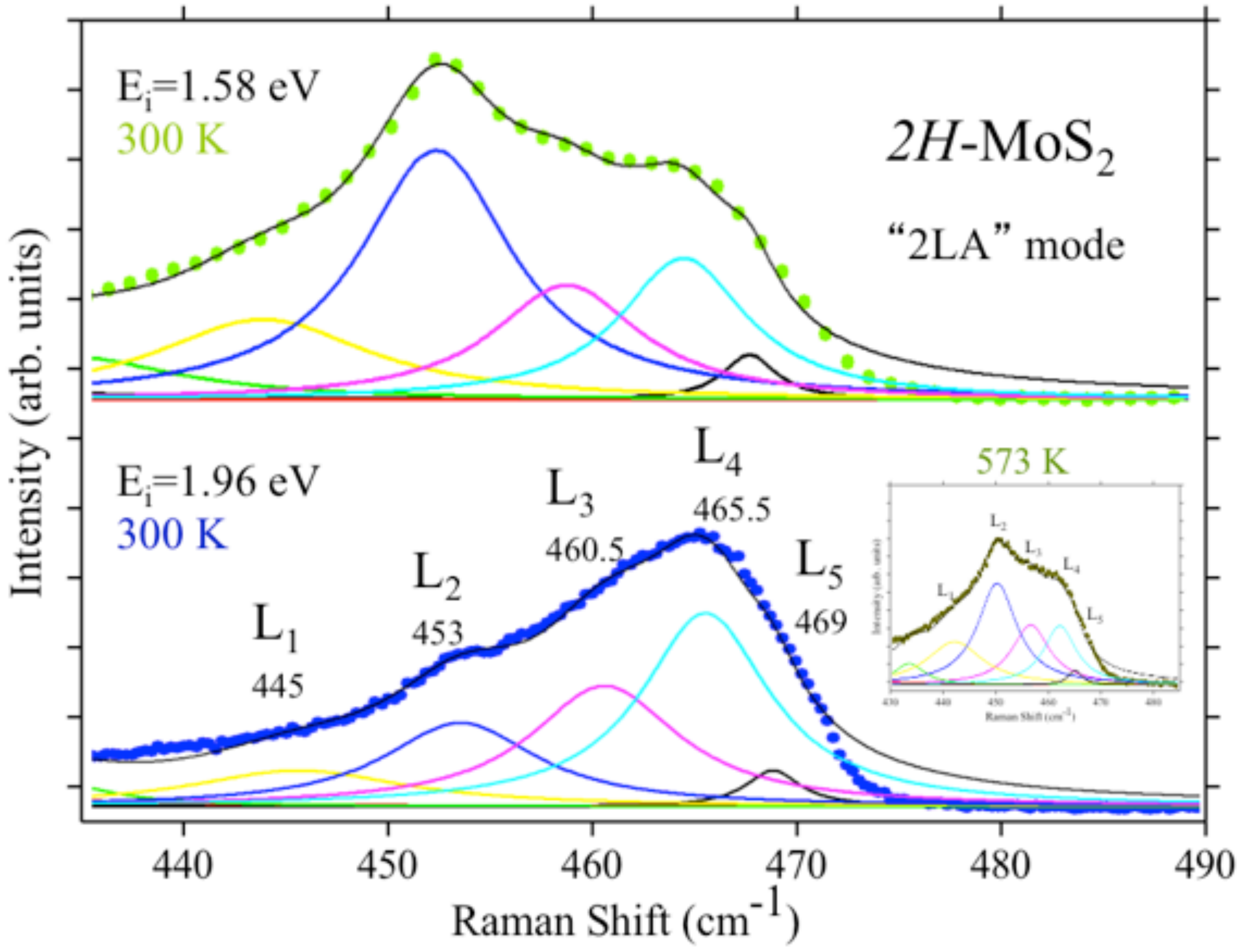

**Figure 7** Lorentzian line fit analysis of the *2H*-$MoS_2$ ~ 460 $cm^{-1}$ band measured at room temperature at $E_i$ = 1.58 eV (top) and $E_i$ = 1.96 eV (bottom). It is evident that the spectra are comprised of at least five contributions [31], denoted in the figure as $L_1$-$L_5$. In the inset of the bottom spectrum shown the spectrum for 573 K with $E_i$ = 1.96 eV. The resemblance of the spectral profile to that of the top spectrum, taken at non-resonant conditions, is evident.

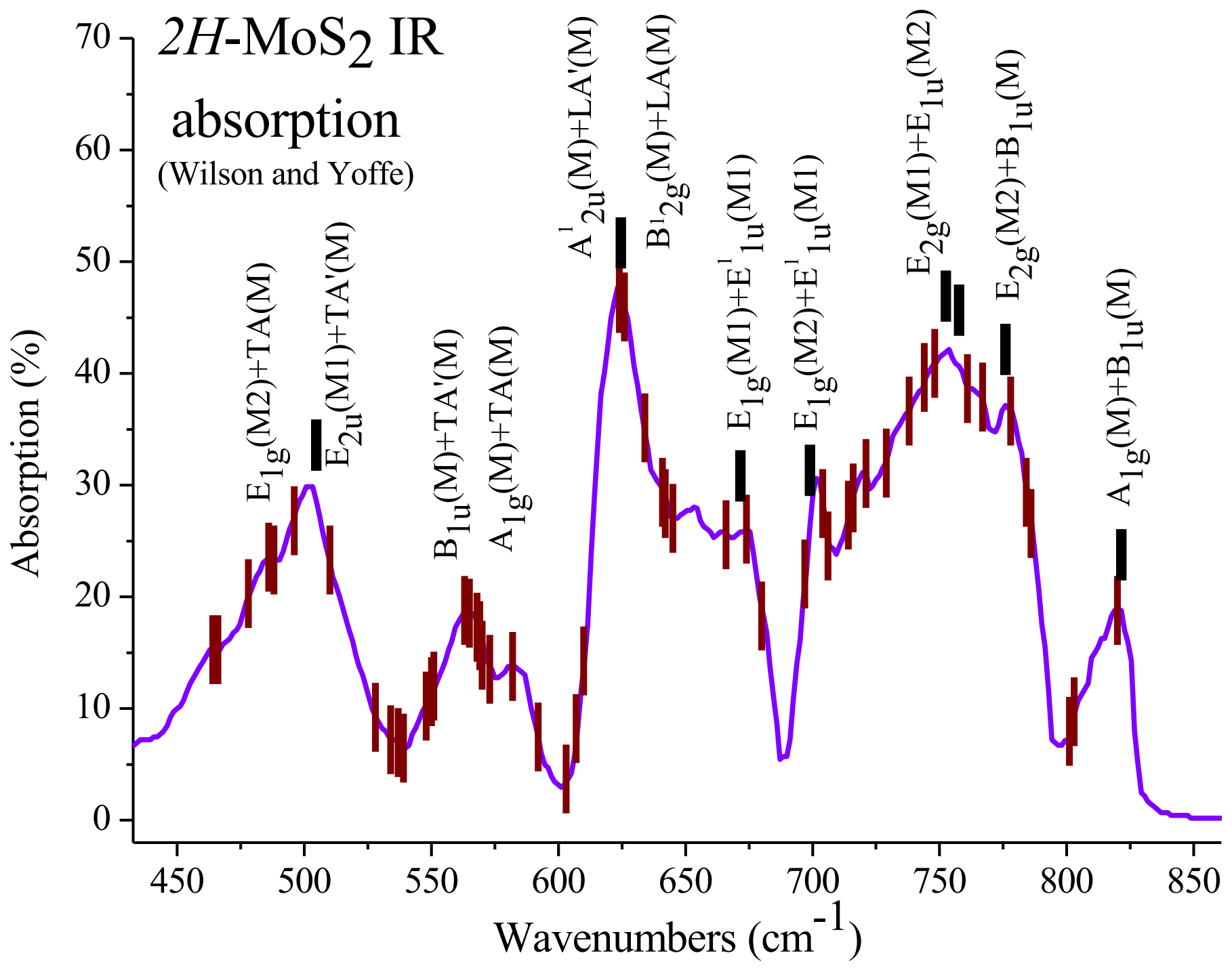

**Figure 8** a) The IR absorption spectrum from a 180 μm layer of *2H*-$MoS_2$ adapted from Willson and Yoffe [18] with the denoted M point IR allowed combinations presented in Table 3. In black stripes, the absorption bands from Agnihorti [32] are indicated after subtracting from them 5 $cm^{-1}$ in order to take into account temperature effects (as they were measured in 77 K). Attached are proposed assignments to some of the bands. K point second-order potential contributors are not included.

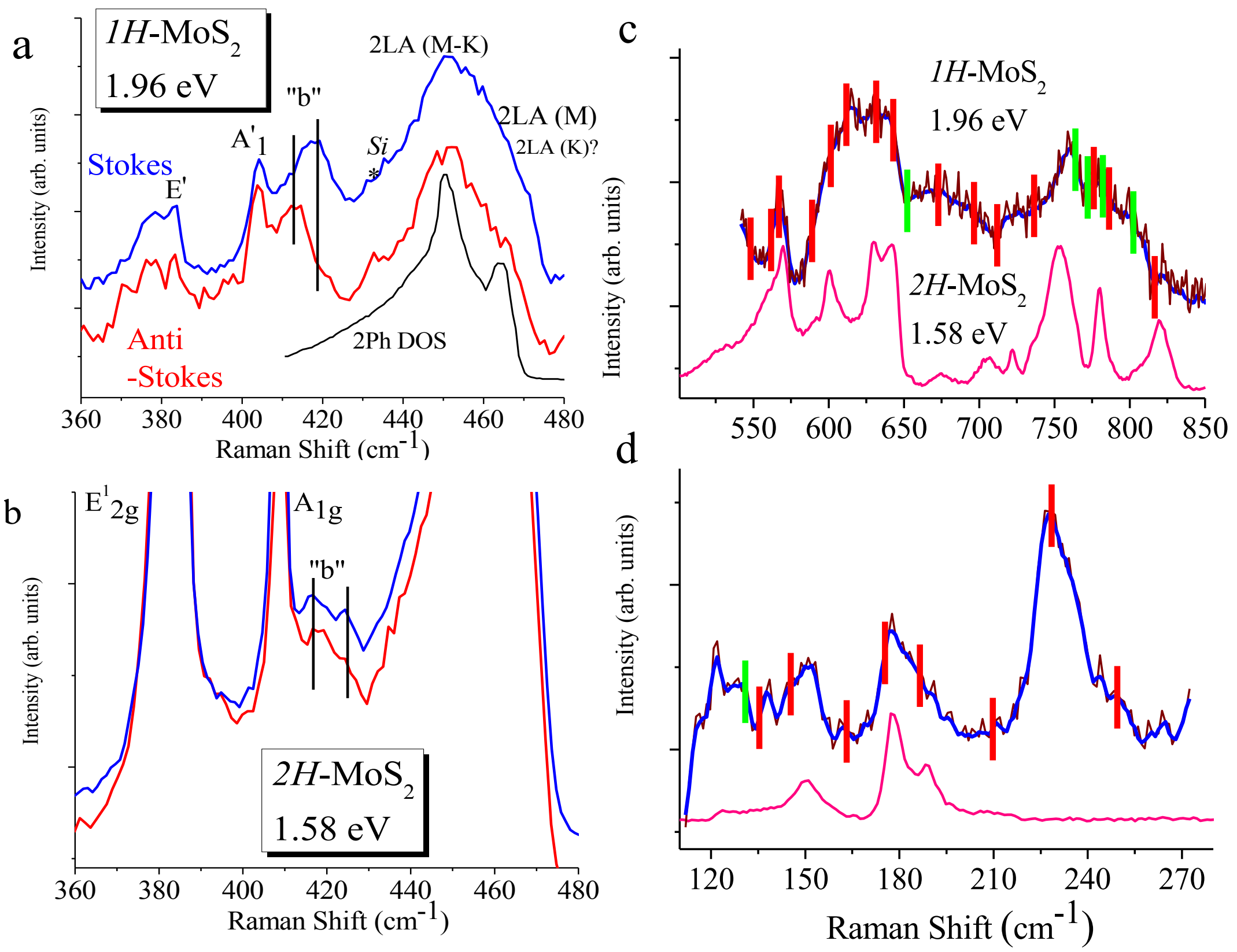

**Figure 9** a) The Stokes (blue) and Anti-Stokes (red) Raman spectrum of single layer measured at 300K and $E_i$ of 1.96 eV (shown in absolute frequency and scaled to similar intensities). The spectrum of the '2LA band' is compared to the 2PDOS of single layer calculated by Ataca *et al.* [22]. The ~5 $cm^{-1}$ shift between the 'b bands' of the two branches (similarly to the bulk *2H* case) are shown. b) Stokes (blue) and Anti-Stokes (red) Raman spectrum of bulk (shown in absolute frequency and scaled to similar intensities), measured at 300 K and $E_i$ of 1.58 eV. Multiphonon spectra for the monolayer in the range of (c) 500-850 $cm^{-1}$ and (d) 110-280 $cm^{-1}$ are compared with that of 1.58 eV bulk *2H* spectra. Expected positions of the multiphonon $A_1$' (M), E' ($M_1$), LA(M) (red) and Γ point (green) are also shown.

# Supporting Information

## S1. Reduction of second-order representations and compatibility relations

Compatibility relations relate the basis functions (wave functions) in going from one wave vector to another belonging to a different symmetry group [1]. Such a situation, for example, occurs in *2H*-$MoS_2$ when going from $\boldsymbol{k} = 0$ (Γ point with $D_{6h}$ symmetry) to an interior $\boldsymbol{k}$ point such as the M Brillouin zone point with $D_{2h}$ symmetry or the K Brillouin zone point with $D_{3h}$ symmetry. In order to construct Table 2 we first reduced the representation of all the $M_i \times M_j$ (i, j=8) and $K_i \times K_j$ (i, j=6) second-order processes into its irreducible constituents and then construct compatibility tables [2, 3] for $\Gamma \xrightarrow{\Sigma} M$ and $\Gamma \xrightarrow{\Lambda} K$. The results of the reduction process are shown in **Tables S1** and **S2** for M and K points, respectively, and the compatibility relations are shown in **Tables S3** and **S4** for M and K points, respectively. For completeness of presentation and further analysis we also show the compatibility table for and $M \xrightarrow{T} K$ in **Table S5**.

After relating the compatibility relations with the reduction process we show in **Tables S6** and **S7** the correlation of M and K points with the Γ point and denote the Raman allowed representations in bold. Let us explore, for example, the $A_g$x$A_g$ and $A_u$x$A_g$ processes in Table S1. In the former, $A_{1g}$ (Γ)+$E_{2g}$ (Γ) comes from the $D_{6h}$ point group at Γ and no additional representation originates from $A_g$ (M)+$B_{1g}$ (M) (see Table S6). This gives a total of $A_{1g}$ (Γ)+$E_{2g}$ (Γ). For $A_u$x$A_g$ none of the symmetries contribute and therefore it is not Raman-active. The contribution of $A_{2u}$ (Γ) indicates that it does have IR activity.

The same arguments apply to *1H*-MoS2 for which the results of the reduction process are shown in **Tables S8.**

## S2. Some important aspects relating the construction of Table 1

Sourisseau *et al.* presented the calculated phonon dispersion of *2H*-$MoS_2$ and phonon symmetries at the M and K points [4]. Recently, more detailed calculated dispersion curves were also published (for example [5]) without specifying the phonon symmetries at the M and K points. The starting point in our analysis (Table 1) is the phonon symmetries assigned in Ref. 4. However, detailed examination of recent calculated dispersion curves (see Fig. 2 of Ref. 5) points to the need to reconsider at least some of the assignments. For example, it seems that some pairs of singly degenerate M point phonons, which remain singly degenerate in Ref. 4, join at the K point to form doubly degenerate phonons. Therefore, we looked at each of those 'suspected' pairs with the aid of the compatibility relations along $\Gamma \overset{\Sigma}{\rightarrow} M$, $\Gamma \overset{\Lambda}{\rightarrow} K$ and $\mathrm{M} \overset{\mathrm{T}}{\rightarrow} K$, which are shown below in Tables S3-S5.

Let us explore, as an example, the ZA and ZA' phonons, which are singly degenerate at M, but seem in the calculations [5] to joint at the K point into a doubly degenerate phonon. The ZA mode starts at Γ with $A_{2u}$ representation (Table 1) and reaches the M point with either $B_{1u}$ or $B_{2g}$ (Table S3) representations (according to Ref. 4, which maintains similar inversion symmetry at Γ and M, it is the former). The ZA' mode starts at Γ with $B_{2g}$ representation and reaches the M point with either $A_u$ or $B_{3g}$ representations (according to Ref. 4 it is the latter). Let us now explore (Table S5) the branch of those phonons from M to K ($\mathrm{M} \overset{\mathrm{T}}{\rightarrow} K$ [3]). The ZA phonon should reach the K point with either $A_2$' or $A_1$'' representations or will combine with another phonon to construct a doubly degenerate E'' phonon at K. The ZA' should also reach the K point with either $A_2$' or $A_1$'' or will combine with another phonon to construct an E'' phonon. Hence, one can construct the following combinations that may lead to ZA and ZA' to combine at K into a doubly degenerate phonons: ($B_{1u}$ (ZA)+$A_u$ (ZA')) and ($B_{2g}$ (ZA)+$B_{3g}$ (ZA'). For this to occur the assignment of one of the bands should be altered from that of Ref. 4.

In order to have a more solid interpretation we need now to explore the compatibility relations of $\Gamma \overset{\Lambda}{\rightarrow} K$ (Table S4): The ZA should lead to $A_2$' or combine with another singly degenerate $A_1$" phonon to construct an E'' phonon at K. The ZA' should also lead to $A_2$' or combine with another singly degenerate $A_1$" phonon to construct an E'' phonon at K.

Since both ZA and ZA' are of $A_2$' representation at the K the scenario of their joining together to form an E" phonon at the K point is not feasible according to Table S4, since for the E'' phonon the irreducible representation contains $\Lambda_3 + \Lambda_2$ (and not $2\Lambda_3$). We therefore argue that singly degenerate ZA(M) and ZA'(M) should evolve into singly degenerate ZA(K) and ZA'(K) with what seems to be [5] very similar frequencies.

After analyzing all the single degenerate phonons at M that may seem to combine into doubly degenerate phonons at K we argue that there may potentially be only two doubly degenerate phonons at K that will be described in the following.

Each of the doubly degenerate modes at the Γ Brillouin zone point generates two phonon branches that reach the M point with single degenerate representations at two different frequencies [3, 4]. From examining the calculated dispersion curves [5] it seems that for the $E^1_{2g}$ (Γ) & $E_{1u}$ (Γ) Davydov doublet the frequencies of $E^1_{2g}$ ($M_1$) and $E_{1u}$ ($M_2$) are ~ 8 $cm^{-1}$ higher than that of $E^1_{2g}$ ($M_2$) and $E_{1u}$ ($M_1$) and for the $E_{1g}$ (Γ) & $E_{2u}$ (Γ) doublet the frequencies of $E_{1g}$($M_1$) and $E_{2u}$($M_1$) are ~24 and ~33 $cm^{-1}$ higher than that of $E_{1g}$($M_2$) and $E_{2u}$ ($M_2$), respectively. While for the latter, the shifts between the two branches were approximated correctly in Ref. 4, for the former it was strongly underestimated with respect to Ref. 5.

We shall explore below the possibility of the construction of doubly degenerate phonon at K from combinations of $E_{1g}$ ($M_1$) + $E_{2u}$ ($M_1$) and $E^1_{2g}$ ($M_2$) + $E_{1u}$($M_1$) phonons. From Table S4 we see that each phonon at M can have one of two representations; even (*g*) or odd (*u*) with respect to inversion. Since, as pointed before, according to Ref. 4 all the phonons at M have the same parity to those of their original phonons at Γ, it is evident from Table S5 that in order for the above scenario to apply the representation of one of the phonons at M of Ref. 4 needs to be altered.

For example, there are two possibilities for $E^1_{2g}$ ($M_1$) to form with $E_{1u}$ ($M_2$) a phonon with E' representation at K (and denoted in Table 1 as $E_{ug}$ ($K_2$)): $B_{1g}$+$A_g$ and $B_{3u}$+$B_{2u}$. In the first case the representation of the $E_{1u}$ ($M_2$) phonon will be altered from $B_{3u}$ to $A_g$ and in the second case the representation of the $E^1_{2g}$ ($M_1$) phonon will be altered from $B_{1g}$ to $B_{3u}$. In our analysis we selected the second option. The reason for this selection is that the $2E^1_{2g}$ ($M_1$) band at 725 $cm^{-1}$, which is Raman allowed at all representations, behaves resonantly (also with respect to temperature dependence, not shown here) but combinations like $E^1_{2g}(M_1)+A_{1g}(M)$ (at 774 $cm^{-1}$) and particularly $E^1_{2g}(M_1)+E^1_{2g}(M_2)$ (at 732 $cm^{-1}$) are definitely absent. We tentatively used this observation as a support for the attribution of the representation of the $E^1_{2g}(M_1)$ phonon to be odd with respect to inversion, with $B_{3u}$ representation rather than $B_{1g}$ (as assigned in Ref. 4). However, since combination bands of $2E^1_{2g}(M_1)$ with all the 'resonant group' phonons are allowed, evidence to their appearance may be found in the spectra. In fact, some of the proposed Raman transition energies in Table 5 may be alternatively assigned to the following combinations with $2E^1_{2g}(M_1)$ ($\phi_1^2$): 312- $\phi_1^2\overline{\chi}$ , 356- $\phi_1^2\overline{\phi_2}$, 566- $\overline{\varphi_1}\phi_1^2$ ,958- $\varphi_2\phi_1^2$,1043- $\varphi_1^2\phi_1^2$, 1117- $\varphi_1\varphi_2\phi_1^2$ (see Table 1 for the various notations)

Finally, it noteworthy that if we would have selected $E_{1u}(M_2)$ phonon to be altered from $B_{3u}$ to $A_g$ the activity and polarization dependent properties of the second-order transitions of the $E_{1u}(M_2)$ and $E^1_{2g}(M_1)$ phonons, as expressed in Table 3, would have been modified. Therefore, with all the 'tools' provided in this paper, Table 3 may be revisited if new theoretical evidence with regard to the symmetry assignments of the Brillouin edge phonons will become available. Similar rationale to the construction of $E_{ug}(K_2)$ applies to the construction of $E_{ug}(K_1)$, which originates from $E_{1g}(M_1)$ and $E_{2u}(M_2)$.

The same procedure that was described above for bulk *2H*-$MoS_2$ was applied to the construction of Tables 6-8 for the monolayer *1H*-$MoS_2$. The compatibility relations and the correlation of M point with the Γ point representations, are shown in **Table S9** and **S10**, respectively.

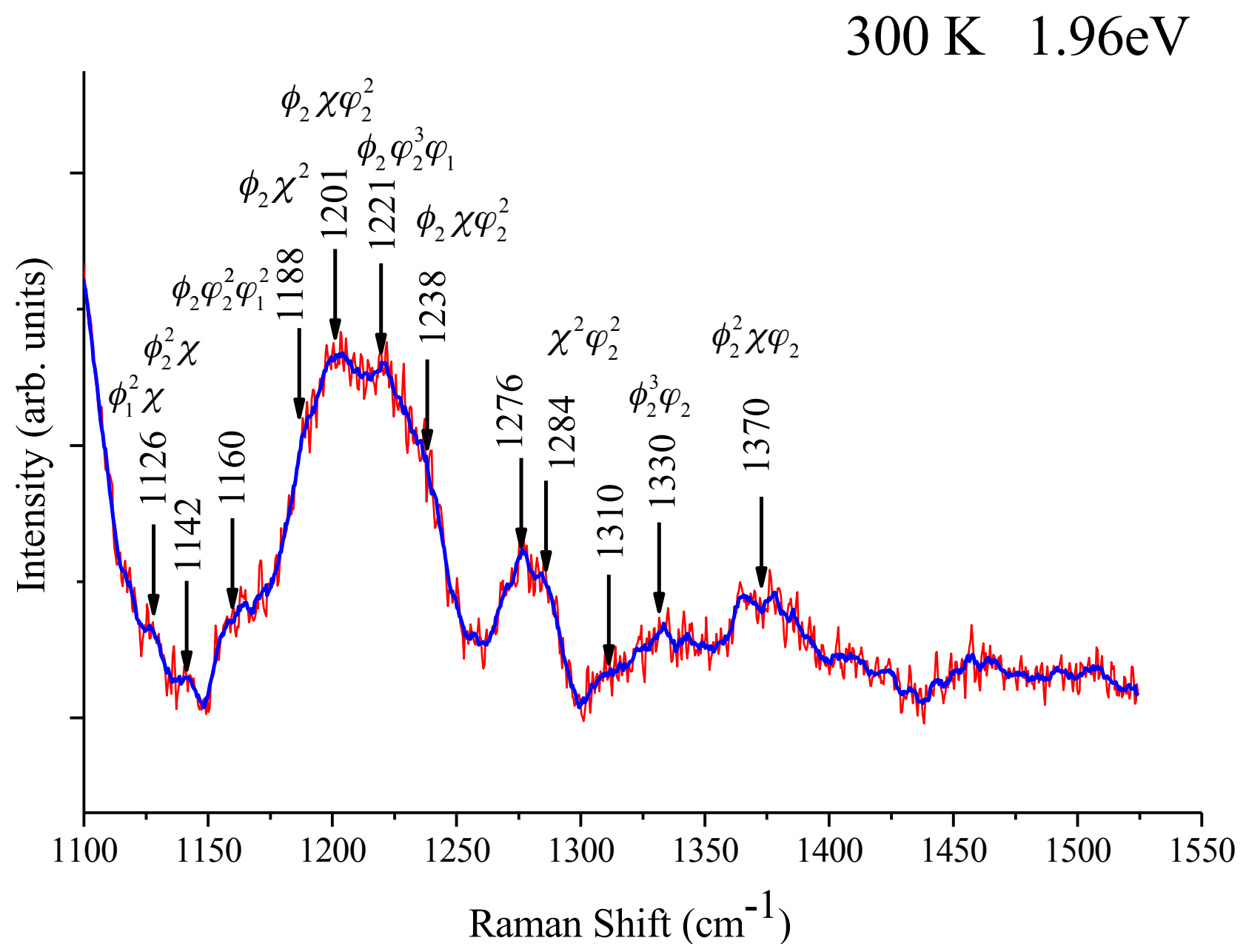

**Figure S1** The 'high end' of the *2H*-$MoS_2$ Raman spectrum, measured at 300 K for $E_i$= 1.96 eV. Assignments of up to 5th order transitions are proposed. For phonons notations see Table 1.

**Table S1** The reduction of the $M_i \times M_j$ (i, j=8) second-order representations to the irreducible representations in *2H*-$MoS_2$. The repesentaions for Γ are denoted in blue and the repesentation for M are denoted in black.

| M/$D_{2h}$ | $A_g$ | $A_u$ | $B_{1g}$ | $B_{1u}$ | $B_{2g}$ | $B_{2u}$ | $B_{3g}$ | $B_{3u}$ |
|---|---|---|---|---|---|---|---|---|
| $A_g$ | $A_{1g}+E_{2g}$ $+A_g+B_{1g}$ | $A_{1u}+E_{2u}$ $+A_u+B_{1u}$ | $A_{2g}+E_{2g}$ $+A_g+B_{1g}$ | $A_{2u}+E_{2u}$ $+A_u+B_{1u}$ | $B_{1g}+E_{1g}+$ $B_{2g}+B_{3g}$ | $B_{1u}+E_{1u}+$ $B_{2u}+B_{3u}$ | $B_{2g}+E_{1g}+$ $B_{2g}+B_{3g}$ | $B_{2u}+E_{1u}+$ $B_{2u}+B_{3u}$ |
| $A_u$ | | $A_{1g}+E_{2g}$ $+A_g+B_{1g}$ | $A_{2u}+E_{2u}$ $+A_u+B_{1u}$ | $A_{2g}+E_{2g}$ $+A_g+B_{1g}$ | $B_{1u}+E_{1u}+$ $B_{2u}+B_{3u}$ | $B_{1g}+E_{1g}+$ $B_{2g}+B_{3g}$ | $B_{2u}+E_{1u}+$ $B_{2u}+B_{3u}$ | $B_{2g}+E_{1g}+$ $B_{2g}+B_{3g}$ |
| $B_{1g}$ | | | $A_{1g}+E_{2g}$ $+A_g+B_{1g}$ | $A_{1u}+E_{2u}$ $+A_u+B_{1u}$ | $B_{2g}+E_{1g}+$ $B_{2g}+B_{3g}$ | $B_{2u}+E_{1u}+$ $B_{2u}+B_{3u}$ | $B_{1g}+E_{1g}+$ $B_{2g}+B_{3g}$ | $B_{1u}+E_{1u}+$ $B_{2u}+B_{3u}$ |
| $B_{1u}$ | | | | $A_{1g}+E_{2g}$ $+A_g+B_{1g}$ | $B_{2u}+E_{1u}+$ $B_{2u}+B_{3u}$ | $B_{2g}+E_{1g}+$ $B_{2g}+B_{3g}$ | $B_{1u}+E_{1u}+$ $B_{2u}+B_{3u}$ | $B_{1g}+E_{1g}+$ $B_{2g}+B_{3g}$ |
| $B_{2g}$ | | | | | $A_{1g}+E_{2g}+$ $A_g+B_{1g}$ | $A_{1u}+E_{2u}+$ $A_u+B_{1u}$ | $A_{2g}+E_{2g}+$ $A_g+B_{1g}$ | $A_{2u}+E_{2u}+$ $A_u+B_{1u}$ |
| $B_{2u}$ | | | | | | $A_{1g}+E_{2g}+$ $A_g+B_{1g}$ | $A_{2u}+E_{2u}+$ $A_u+B_{1u}$ | $A_{2g}+E_{2g}+$ $A_g+B_{1g}$ |
| $B_{3g}$ | | | | | | | $A_{1g}+E_{2g}+$ $A_g+B_{1g}$ | $A_{1u}+E_{2u}+$ $A_u+B_{1u}$ |
| $B_{3u}$ | | | | | | | | $A_{1g}+E_{2g}+$ $A_g+B_{1g}$ |

**Table S2** The reduction of the $K_i \times K_j$ (i, j=6) second-order representations to the irreducible representations in *2H*-$MoS_2$. The repesentaion for Γ are denoted in blue and the repesentation for K in black.

| K/$D_{3h}$ | $A'_1$ | $A''_1$ | $A'_2$ | $A''_2$ | $E'$ | $E''$ |
|---|---|---|---|---|---|---|
| $A'_1$ | $A_{1g}+B_{1u}$ $+A'_1$ | $A_{2g}+B_{2u}$ $+A''_1$ | $A_{1u}+B_{1g}$ $+A'_2$ | $A_{2u}+B_{2g}$ $+A''_2$ | $E_{2g}+E_{1u}$ $+E'$ | $E_{2u}+E_{1g}$ $+E''$ |
| $A''_1$ | | $A_{1g}+B_{1u}$ $+A'_1$ | $A_{2u}+B_{2g}$ $+A''_2$ | $A_{1u}+B_{1g}$ $+A'_2$ | $E_{2g}+E_{1u}$ $+E'$ | $E_{2u}+E_{1g}$ $+E''$ |
| $A'_2$ | | | $A_{1g}+B_{1u}$ $+A'_1$ | $A_{2g}+B_{2u}$ $+A''_1$ | $E_{2u}+E_{1g}$ $+E''$ | $E_{2g}+E_{1u}$ $+E'$ |
| $A''_2$ | | | | $A_{1g}+B_{1u}$ $+A'_1$ | $E_{2u}+E_{1g}$ $+E''$ | $E_{2g}+E_{1u}$ $+E'$ |
| $E'$ | | | | | $A_{1g}+A_{2g}+B_{1u}$ $+B_{2u}+E_{2g}+E_{1u}$ $+A'_1+A''_1+E'$ | $A_{1u}+A_{2u}+B_{1g}+$ $B_{2g}+E_{2u}+E_{1g}$ $+A'_2+A''_2+E''$ |
| $E''$ | | | | | | $A_{1g}+A_{2g}+B_{1u}+$ $B_{2u}+E_{2g}+E_{1u}$ $+A'_1+A''_1+E'$ |

**Table S3** Compatability relations along Σ in *2H*-$MoS_2$

| Compatibility relation between M and Σ | | Compatibility relation between Γ and Σ | |
|---|---|---|---|
| $D_{2h}(M)$ | Irreducible representations | $D_{6h}(\Gamma)$ | Irreducible representations |
| $A_g$ | $\Sigma_1$ | $A_{1g}$ | $\Sigma_1$ |
| $A_u$ | $\Sigma_4$ | $A_{1u}$ | $\Sigma_4$ |
| $B_{1g}$ | $\Sigma_2$ | $A_{2g}$ | $\Sigma_2$ |
| $B_{1u}$ | $\Sigma_3$ | $A_{2u}$ | $\Sigma_3$ |
| $B_{2g}$ | $\Sigma_3$ | $B_{1g}$ | $\Sigma_3$ |
| $B_{2u}$ | $\Sigma_2$ | $B_{1u}$ | $\Sigma_2$ |
| $B_{3g}$ | $\Sigma_4$ | $B_{2g}$ | $\Sigma_4$ |
| $B_{3u}$ | $\Sigma_1$ | $B_{2u}$ | $\Sigma_1$ |
| | | $E_{1g}$ | $\Sigma_4 + \Sigma_3$ |
| | | $E_{1u}$ | $\Sigma_2 + \Sigma_1$ |
| | | $E_{2g}$ | $\Sigma_2 + \Sigma_1$ |
| | | $E_{2u}$ | $\Sigma_4 + \Sigma_3$ |

**Table S4** Compatability relations along Λ in *2H*-$MoS_2$

| Compatibility relation between K and Λ | | Compatibility relation between Γ and Λ | |
|---|---|---|---|
| $D_{3h}(K)$ | Irreducible representations | $D_{6h}(\Gamma)$ | Irreducible representations |
| $A_1'$ | $\Lambda_1$ | $A_{1g}$ | $\Lambda_1$ |
| $A_1''$ | $\Lambda_2$ | $A_{1u}$ | $\Lambda_2$ |
| $A_2'$ | $\Lambda_3$ | $A_{2g}$ | $\Lambda_4$ |
| $A_2''$ | $\Lambda_4$ | $A_{2u}$ | $\Lambda_3$ |
| $E'$ | $\Lambda_4 + \Lambda_1$ | $B_{1g}$ | $\Lambda_2$ |
| $E''$ | $\Lambda_3 + \Lambda_2$ | $B_{1u}$ | $\Lambda_1$ |
| | | $B_{2g}$ | $\Lambda_3$ |
| | | $B_{2u}$ | $\Lambda_4$ |
| | | $E_{1g}$ | $\Lambda_3 + \Lambda_2$ |
| | | $E_{1u}$ | $\Lambda_4 + \Lambda_1$ |
| | | $E_{2g}$ | $\Lambda_4 + \Lambda_1$ |
| | | $E_{2u}$ | $\Lambda_3 + \Lambda_2$ |

**Table S5** Compatability relations along T in *2H*-$MoS_2$

| Compatibility relation between K and T | | Compatibility relation between M and T | |
|---|---|---|---|
| $D_{3h}(K)$ | Irreducible representations | $D_{2h}(M)$ | Irreducible representations |
| $A_1'$ | $T_1$ | $A_g$ | $T_1$ |
| $A_1''$ | $T_2$ | $A_u$ | $T_2$ |
| $A_2'$ | $T_3$ | $B_{1g}$ | $T_4$ |
| $A_2''$ | $T_4$ | $B_{1u}$ | $T_3$ |
| $E'$ | $T_1 + T_4$ | $B_{2g}$ | $T_2$ |
| $E''$ | $T_2 + T_3$ | $B_{2u}$ | $T_1$ |
| | | $B_{3g}$ | $T_3$ |
| | | $B_{3u}$ | $T_4$ |

**Table S6** The correlation of the M point representations with the Γ point in *2H*-$MoS_2$

| |
|---|
| $A_g(M)$ = **$A_{1g}(\Gamma)$+ $E_{2g}(\Gamma)$** |
| $A_u(M)$ = $A_{1u}(\Gamma)$+$E_{2u}(\Gamma)$ |
| $B_{1g}(M)$ = $A_{2g}(\Gamma)$+**$E_{2g}(\Gamma)$** |
| $B_{1u}(M)$ = **$A_{2u}(\Gamma)$**+$E_{2u}(\Gamma)$ |
| $B_{2g}(M)$ = $B_{1g}(\Gamma)$+**$E_{1g}(\Gamma)$** |
| $B_{2u}(M)$ = $B_{1u}(\Gamma)$+**$E_{1u}(\Gamma)$** |
| $B_{3g}(M)$ = $B_{2g}(\Gamma)$**+ $E_{1g}(\Gamma)$** |
| $B_{3u}(M)$ = $B_{2u}(\Gamma)$**+ $E_{1u}(\Gamma)$** |

**Table S7** The correlation of the K point representations with the Γ point in *2H*-$MoS_2$

| |
|---|
| $A_1'(K)$ **= $A_{1g}(\Gamma)$**+$B_{1u}(\Gamma)$**+$E_{2g}(\Gamma)$**+**$E_{1u}(\Gamma)$** |
| $A_1''(K)$ = $A_{1u}(\Gamma)$+$B_{1g}(\Gamma)$+**$E_{1g}(\Gamma)$+**$E_{2u}(\Gamma)$ |
| $A_2'(K)$ **= $A_{2u}(\Gamma)$**+$B_{2g}(\Gamma)$+**$E_{1g}(\Gamma)$+**$E_{2u}(\Gamma)$ |
| $A_2''(K)$ = $A_{2g}(\Gamma)$+$B_{2u}(\Gamma)$+**$E_{2g}(\Gamma)$+$E_{1u}(\Gamma)$** |
| $E'(K)$ **= $E_{2g}(\Gamma)$+$E_{1u}(\Gamma)$** |
| $E''(K)$ = $E_{2u}(\Gamma)$**+$E_{1g}(\Gamma)$** |

**Table S8** The reduction of the $M_i \times M_j$ (i, j=4) second-order representations to the irreducible representations in *1H*-$MoS_2$. The repesentaions for Γ are denoted in blue and the repesentation for M are denoted in black.

| M/$C_{2v}$ | $A_1$ | $A_2$ | $B_1$ | $B_2$ |
|---|---|---|---|---|
| $A_1$ | $A_1'+E'$ $+A_1+B_1$ | $A_1''+E''$ $+A_2+B_2$ | $A_2'+E'$ $+A_1+B_1$ | $A_2''+E''$ $+A_2+B_2$ |
| $A_2$ | | $A_1'+E'$ $+A_1+B_1$ | $A_2''+E''$ $+A_2+B_2$ | $A_2'+E'$ $+A_1+B_1$ |
| $B_1$ | | | $A_1'+E'$ $+A_1+B_1$ | $A_1''+E''$ $+A_2+B_2$ |
| $B_2$ | | | | $A_1'+E'$ $+A_1+B_1$ |

**Table S9** Compatability relations along Σ in *1H*-$MoS_2$

| Compatibility relation between M and Σ | | Compatibility relation between Γ and Σ | |
|---|---|---|---|
| $C_{2v}$(M) | Irreducible representations | $D_{3h}$(Γ) | Irreducible representations |
| $A_1$ | $\Sigma_1$ | $A_1'$ | $\Sigma_1$ |
| $A_2$ | $\Sigma_3$ | $A_1''$ | $\Sigma_3$ |
| $B_1$ | $\Sigma_4$ | $A_2'$ | $\Sigma_2$ |
| $B_2$ | $\Sigma_2$ | $A_2''$ | $\Sigma_4$ |
| | | $E'$ | $\Sigma_2 + \Sigma_1$ |
| | | $E''$ | $\Sigma_4 + \Sigma_3$ |

**Table S10** The correlation of the M point representations with the Γ point in *1H*-$MoS_2$

| |
|---|
| $A_1$(M) = $\mathbf{A_1'(\Gamma)}$+ $\mathbf{E'(\Gamma)}$ |
| $A_2$ (M) = $A_1''(\Gamma)$+ $\mathbf{E''(\Gamma)}$ |
| $B_1$ (M) = $A_2''(\Gamma)$+ $\mathbf{E''(\Gamma)}$ |
| $B_2$ (M) = $A_2'(\Gamma)$+ $\mathbf{E'(\Gamma)}$ |